\documentclass{article}

\usepackage{amssymb}
\usepackage{wrapfig}

\usepackage{wrapfig}
\usepackage{mathrsfs}
\usepackage{amsbsy,cancel}
\usepackage{color}
\usepackage{graphicx}

\def \Sscr{\mathcal S}
\def\sign{\mathrm sign}
\def \ve{\varepsilon}
\usepackage{verbatim}
\def\wt{\widetilde}

\def\wh{\widehat}

\def\ds{\displaystyle}

\renewcommand{\theequation}{\arabic{section}-\arabic{equation}}
\makeatletter
\@addtoreset{equation}{section}
\makeatother

\def\le{\left}
\def\ri{\right}
\def\QED{{\bf Q.E.D.}\par\vskip 5pt}

\def\bc{\begin{corollary}}
\def\ec{\end{corollary}}
\def\&{&{\hskip -20pt}}
\def\m{\mathop}

\def \s{\mathfrak s}
\def\ov{\overline}
\def\br{\begin{remark}\rm\small}
\def\1{{\bf 1}}
\def\er{\end{remark}}
\def\bt{\begin{theorem}}
\def\et{\end{theorem}}

\def\bx{\begin{example}}
\def\ex{\end{example}}
\def\bi{\begin{itemize}}
\def\ei{\end{itemize}}
\def\bd{\begin{definition}}
\def\ed{\end{definition}}
\def\bp{\begin{proposition}\rm}
\def\bl{\begin{lemma}\em}
\def\el{\end{lemma}}
\def\ep{\end{proposition}}
\def\bea{\begin{eqnarray}}
\def\eea{\end{eqnarray}}
\def \pa{\partial}
\def\C{{\mathbb C}}
\def\R{{\mathbb R}}
\def\N{{\mathbb N}}
\def\Z{{\mathbb Z}}

\newenvironment{bmatrix}
{\le[\begin{array}{cc}} {\end{array}\ri]}
\renewenvironment{pmatrix}
{\le(\begin{array}{cc}} {\end{array}\ri)}
\newtheorem{problem}{Problem}[section]
\newtheorem{theorem}{Theorem}[section]
\newtheorem{example}{Example}[section]
\newtheorem{coroll}{Corollary}[section]
\newtheorem{examps}{Examples}[section]

\newtheorem{lemma}{Lemma}[section]
\newtheorem{remark}{Remark}[section]
\newtheorem{remarks}[remark]{Remarks}
\newtheorem{proposition}{Proposition}[section] 
\newtheorem{definition}{Definition}[section]
\def\br{\begin{remark}}
\def\er{\end{remark}}
\def\bt{\begin{theorem}}
\def\et{\end{theorem}}
\def\bc{\begin{coroll}}
\def\ec{\end{coroll}}
\def\brs{\begin{remarks} \rm\
\begin{enumerate}}
\def\ers{\end{enumerate}\end{remarks}}
\def\bl{\begin{lemma}}
\def\el{\end{lemma}}
\def\bxs{\begin{examps}. \rm\begin{enumerate}}
\def\exs{\end{enumerate}\end{examps}}
\def\bd{\begin{definition}}
\def\ed{\end{definition}}
\def\bp{\begin{proposition}}
\def\ep{\end{proposition}}
\def\be{\begin{equation}}
\def\ee{\end{equation}}

\def\d{{\rm d}}
\def\bea{\begin{eqnarray}}
\def\eea{\end{eqnarray}}
\def\beas{\begin{eqnarray*}}
\def\eeas{\end{eqnarray*}}
\def\gt{\hat\gamma}
\def \hf{\frac{1}{2}}
\def \pa{\partial}

\def \ra{\rightarrow}

\def\C{{\mathbb C}}
\def \D{\mathbb D}
\def \A{\mathbf A}
\def\a{\alpha}
\def\g{\gamma}
\def\k{\varkappa}
\def\l{\lambda}
\def\m{\mu}
\def\s{\sigma}
\def\t{\tau}
\def\z{\zeta}
\def \B{\mathbf B}

\def\R{{\mathbb R}}
\def\N{{\mathbb N}}

\def\Z{{\mathbb Z}}

\date{}
\begin{document}
%

\baselineskip 16pt plus 1pt minus 1pt
\begin{titlepage}
\begin{flushright}
\end{flushright}
\vspace{0.2cm}
\begin{center}
\begin{Large}
\textbf{Universality in the profile of the semiclassical limit solutions to the focusing Nonlinear Schr\"odinger equation at the first breaking curve}
\end{Large}\\
\bigskip
{\large M. Bertola}$^{\dagger\ddagger}$\footnote{Work supported in part by the Natural
   Sciences and Engineering Research Council of Canada (NSERC)}\footnote{bertola@crm.umontreal.ca},  
A. Tovbis$^{\sharp}$ 
\\
\bigskip
\begin{small}
$^{\dagger}$ {\em Centre de recherches math\'ematiques,
Universit\'e de Montr\'eal\\ C.~P.~6128, succ. centre ville, Montr\'eal,
Qu\'ebec, Canada H3C 3J7} \\
\smallskip
$^{\ddagger}$ {\em  Department of Mathematics and
Statistics, Concordia University\\ 1455 de Maisonneuve W., Montr\'eal, Qu\'ebec,
Canada H3G 1M8} \\
\smallskip
$^{\sharp}$ {\em  University of Central Florida
	Department of Mathematics\\
	4000 Central Florida Blvd.
	P.O. Box 161364
	Orlando, FL 32816-1364
} \\
\end{small}
\end{center}
\bigskip
\begin{center}{\bf Abstract}\\
\end{center}
{We consider  the {semiclassical (zero-dispersion) limit of the  one-dimensional focusing Nonlinear  Schr\"odinger equation (NLS) with decaying potentials.
If a potential is a simple rapidly oscillating wave (the period has the order of the semiclassical parameter $\ve$) with  modulated amplitude and phase, 
t}he space-time plane subdivides into regions {of} qualitatively different behavior, with the boundary between them consisting typically of collection of 
piecewise smooth arcs ({\em breaking curve(s)}). { In the first region the evolution of the potential 
is ruled by modulation equations (Whitham equations), but for every value of the space variable  $x$ there is a moment of transition (breaking),
where the solution develops fast, quasi-periodic behavior, i.e.,   the amplitude becomes also fastly oscillating at scales of order $\varepsilon$.
The very first point of  such transition is called the point of  {\em  gradient catastrophe}.}
We study the detailed asymptotic behavior of the left and right edges 
of the interface between these two regions at any time after the gradient catastrophe. The main finding is that the first oscillations in the amplitude are 
of nonzero asymptotic size even as $\varepsilon$ tends to zero, and they display two separate natural scales; of order $\mathcal O(\varepsilon)$ in the 
parallel direction to the breaking curve in the $(x,t)$-plane, and of order $\mathcal O(\varepsilon \ln \varepsilon)$ in a transversal direction. 
The study is based upon the inverse-scattering method and the nonlinear steepest descent method.
}
\medskip
\bigskip
\bigskip
\bigskip
\bigskip

\end{titlepage}
\tableofcontents

\section{Introduction and main results}
\label{sectintro}
The focusing Nonlinear Schr\"odinger (NLS) equation,
\be  \label{FNLS}
i\varepsilon\partial_t q + \hf\varepsilon^2\partial_x^2 q + |q|^2q=0, \ \ \ \ \ \
\ee
where $x\in\R$ and $t\ge 0$ are space-time variables,
is  a basic model for self-focusing and
self-modulation  e.g. it governs nonlinear transmission in optical fibers; 
it can also be derived as a  modulation equation for  general nonlinear systems. 
It was first integrated by Zakharov and Shabat \cite{ZS}
who produced a Lax pair for it and used the inverse scattering
procedure to describe general decaying solutions ($\lim_{|x|\to 0}q(x,0)=0$) in terms of radiation  and solitons. Throughout this work, we will
use the abbreviation NLS to mean ``focusing Nonlinear Schr\"odinger equation".

Our  interest in the  semiclassical
limit ($\varepsilon \to 0$) of NLS  stems largely from  its
{\em modulationally unstable} behavior, as shown  by  Forest 
and Lee \cite{FL}, the modulation system  for NLS can be expressed as 
a set of nonlinear PDE with  complex characteristics; 
thus, the system  is ill posed as an initial value 
problem with Cauchy data.  As a result,   initial data 
\be \label{IDe}
q(x,0,\varepsilon)=A(x)e^{i\Phi(x)/\varepsilon},
\ee
i.e. a plane wave
with amplitude modulated by $A(x)$ and phase modulated by $\Phi(x)$  
are expected to  break immediately into some other, 
presumably more disordered,  wave form when the functions $A(x)$ and $\Phi(x)$ 
possess no special properties.

\begin{figure}
\includegraphics[width = 0.99\textwidth]{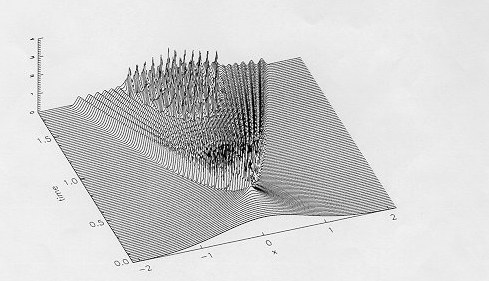}
\caption{Absolute value $|q(x,t,\varepsilon)|$ of a solution  $q(x,t,\varepsilon)$ to the focusing NLS (\ref{FNLS})
versus $x,t$ coordinates from \cite{CMM}. Here $A(x)=e^{-{x^2}}$, $\Phi(x)=-\tanh x$ and $\ve=0.02$.}
\label{Cai}\end{figure}

However, in the case of an {\em analytic} initial data, the NLS evolution  
displays
some  ordered structure 
instead of the disorder suggested by the modulational
instability, as shown by Fig. \ref{Cai}, see \cite{MillerKamvissis},
\cite{CT} and \cite{CMM}. 

Fig.  \ref{Cai} from \cite{CMM} clearly identifies regions {where}  different types of  behavior 
of {the} solution $q(x,t,\varepsilon)$ appear. These regions are separated by some  independent of $\varepsilon$ curves in the $x,t$ plane
that are called {\bf breaking curves} or {\bf nonlinear caustics}. 
Within each region, the strong asymptotics  of $q(x,t,\varepsilon)$ can be expressed in terms of Riemann
Theta-functions to within an  error  term of order   $\mathcal O(\varepsilon)$ that is uniform on compact subsets of the region (\cite{KMM} for the pure soliton case and \cite{TVZ1} for the general case). In this context, regions of different asymptotic behavior of 
$q(x,t,\varepsilon )$ corresponds to the different genera of the hyperelliptic Riemann surface whose Theta-functions enter in the asymptotic description.
No error estimates for the asymptotics on the
breaking curve were studied so far, although it was expected that the
accuracy should be of the 
order $O(\varepsilon^\frac{1}{2})$ at any regular point on the breaking curve. A visual inspection of Fig. \ref{Cai} suggests that the behavior of 
$q$ at the tip of the breaking curve is different from the behavior elsewhere on the breaking curve.

The tip-point is called a point of  {\bf gradient catastrophe}, or {\bf elliptic umbilical
singularity} (\cite{DGK}).) The main goal of this paper is to analyze the leading order asymptotic behavior of
the solution $q(x,t,\varepsilon )$ on and around the breaking curve except a vicinity of the gradient catastrophe
point, {and to obtain the corresponding}  error estimate. More precisely, we will examine a strip around the breaking curve between the genus zero and genus two regions,  
which has the  width of order  $\mathcal O(\varepsilon\ln\varepsilon )$.

\subsection{Description of results}
The main goal of the paper is to provide a detailed asymptotic analysis in the zero-dispersion (or semiclassical) limit ($\varepsilon\to 0$) in the neighborhood of the breaking curve, that is,
the frontier in $(x,t)$-space that separates the region of smooth behavior ({\em genus zero} region) from the region of fast oscillations ({\em genus two} region), which is 
clearly seen on Fig. \ref{Cai}.

The main highlight is an ``universal'' expression for the behavior of the first oscillations as we egress from the genus zero region into the genus two one. Here ``universality'' means 
that the expression does not depend upon the details of the initial data, or rather, it depends on it only through a few parameters that are explicitly computable. 
In order to give a taste of the expression in consideration we need to present a minimum of background notation, with more details contained in the body of the paper.

The zero-dispersion limit of (\ref{FNLS})  can be addressed in terms of modulation equations involving one pair of complex conjugate Riemann invariants, say $\alpha,\ov \alpha$. 
The invariant $\alpha=\alpha(x,t)$ (which we choose in the upper half plane) depends on time/space according to the modulation equations (also called Whitham equations). 
The asymptotic behavior for the solution of the NLS equation in the genus zero region is then provided by \cite{TVZ1}
\bea
q(x,t) \sim -b(x,t) {\rm e}^{\frac i \varepsilon \Phi(x,t)}(1+o(1)),~~~~~~~~{\rm where}\label{zeroappr}\\
\alpha(x,t) = a(x,t) + i b(x,t)\cr
\pa_x \Phi(x,t) =-2 a(x,t)\nonumber
\eea
and $o(1)$  denotes  some infinitesimal quantity in $\varepsilon$. 
However, as we leave the genus-zero region, such approximation breaks down: the term that was infinitesimal $o(1)$ becomes suddenly of finite order and a different approximation scheme must be employed. 
The modulation equations alone cannot detect such sudden change of behavior, since they are equations that are obtained from a formal manipulation. 
The modern approach to the problem requires the introduction of a special function, the {\bf $g$-function}; this is a locally analytic function in the complex spectral  plane  of 
the associated linear problem. 
It is defined in terms of the scattering transform of the initial data: without entering the details we may say that  
in our problem the initial data is  encoded in a Schwarz-symmetrical analytic function $f_0(z)$  ($\Im f_0(z)$  generically has  a jump discontinuity along the real axis). 
Taking into account the time ($t$) evolution and normalization with respect to $x$, one obtains 
\be
f(z):= f(z; x,t) = f_0(z) - x z - 2t z^2\ .
\ee 
The $g$--function then is obtained as solution of a  scalar Riemann--Hilbert problem (RHP) on a free boundary with the jump $f(z)$
\footnote{This means that the contour where the jump is has to be determined implicitly along the way of the construction of the solution itself.}.

For the reader unfamiliar with the method  we may compare the {determination of the}  free-boundary to the determination of the steepest descent contour for Laplace-type integral. The main requirements (but this is a severely shortened list, see Section \ref{sectreview} for the details) is that 
\be
g_+ (z) + g_-(z) = f(z; x,t)\ ,\qquad z\in \gamma_m\label{cond1}
\ee
where $\gamma_m$ is an arc of the said free boundary and the sign in the subscript denotes the left/right boundary value; moreover  $g$ is bounded at  infinity and with some growth requirements at the endpoints (if any) of $\gamma_m$. The function $g(z)=g(z;x,t)$, defined through the above mentioned RHP,
must be supplemented by several inequalities that involve  
\be\label{hgf}
h(z;x,t):= 2g(z;x,t) - f(z;x,t)\ ,
\ee
and the {\bf sign distribution} of its {\bf imaginary part} $\Im (h)$. 
All the functions $f,g,h$ are Schwartz-symmetrical, in particular $h(z;x,t) = \ov {h(\ov z; x,t)}$; this means that the sign-distribution requirements in the lower half-plane are the opposite of those in the upper half, $\C_+$. 

One of the main requirements is that $\Im (h)$ in $\C_+$ must be {\bf positive} along the complement $\g_c$ 
of $\gamma_m$ within the above-mentioned free-boundary, { and must be negative on both sides of 
$\g_m$, the latter condition implying  that $\Im h=0$ on $\g_m$}. 
What happens is that one of these particular requirements fails in certain regions of the $(x,t)$ plane; this is where the zero-dispersion asymptotic behavior of the solution changes its behavior. The typical {\em phase portrait} is depicted in Fig. \ref{phasediagram}.  The boundary between the two regions consists of points $(x,t)$, {for which  $\Im(h(z;x,t)=0$ and simultaneously $h'(z;x,t)=0$ at some point $z$ of the spectral plane; we denote by $\eta(x,t)$ the value of $z\in \mathbb C_+$ satisfying $h'(z;x,t)=0$ for all $x,t$ in a vicinity of the breaking curve.} 
Because of the analyticity of $h$ this point $\eta$ must be a saddle--point for $\Im h$ and $h$ can be written as 
\be
h(z; x,t) = S(x,t) - \frac {C(x,t)}{2i} (z-\eta)^2 + \mathcal O( (z-\eta)^3)\label{SC}\ .
\ee
(the factors are just for convenience). Note that the breaking curve is now given by $\Im S(x,t)=0$. Generically, the second derivative coefficient $C(x,t)$ does not vanish: this corresponds to the smooth part of the boundary between the two regions in the sketch of Fig. \ref{phasediagram}. The angular point of the boundary (where $C(x_0,t_0)=0$) is the so--called {\bf point of umbilical gradient catastrophe} and will be dealt with in a forthcoming publication.

The critical value $S(x,t)\in \C$ plays the first major role in our story: indeed we have 
\begin{itemize}
\item The function $S(x,t)$ is a local diffeomorphism for $x\in \R$, $t\in \R_+$ with values in $\C\sim \R^2$ except at the point of gradient catastrophe (Lemma \ref{Jacobian}).
\item The function $S(x,t)$ maps a tubular neighborhood of the breaking curve not containing the point of gradient catastrophe diffeomorphically on a tubular neighborhood of the real $S$--axis (Thm. \ref{breakingcurve}).
\end{itemize}
This means that we can use $\Im S$ as a parametrization in a transversal direction to the breaking curve, while using $\Re S$ as a parametrization in the parallel direction. 
The paper focuses on the analysis of such a tubular neighborhood with a width of order $\mathcal O(\ve\ln \ve)$; more precisely we introduce the {\bf exploration parameter} $\varkappa(x,t)$  (Def. \ref{explore}) by 
\be
S(x,t)= \frac {\vartheta(x,t)} 2 +  \frac i 2 \varkappa(x,t) \ve\ln \ve\ .  
\ee
The parameter $\varkappa$ is positive on the left breaking curve (negative on the right one, respectively) as we enter the oscillatory zone (in wavy lines in Fig. \ref{phasediagram}); a finite value of $\varkappa$ corresponds to an infinitesimal {\em exploration}.

\begin{figure}[htbp]
\begin{center}
\includegraphics[width=0.5\textwidth]{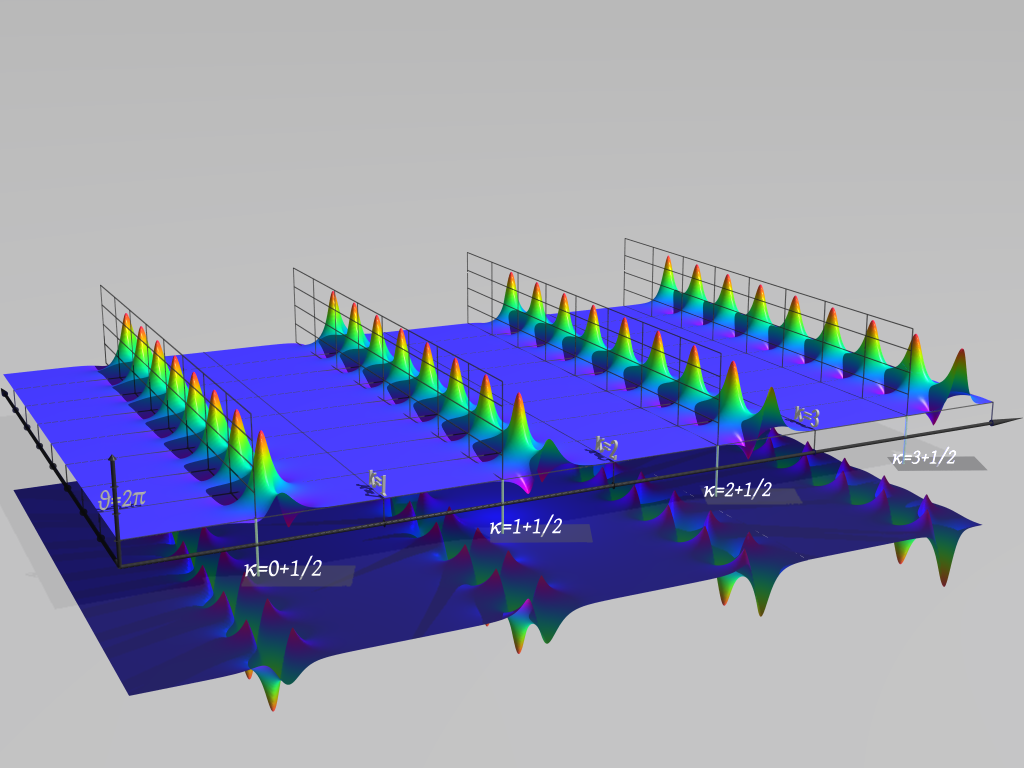}\includegraphics[width=0.5\textwidth]{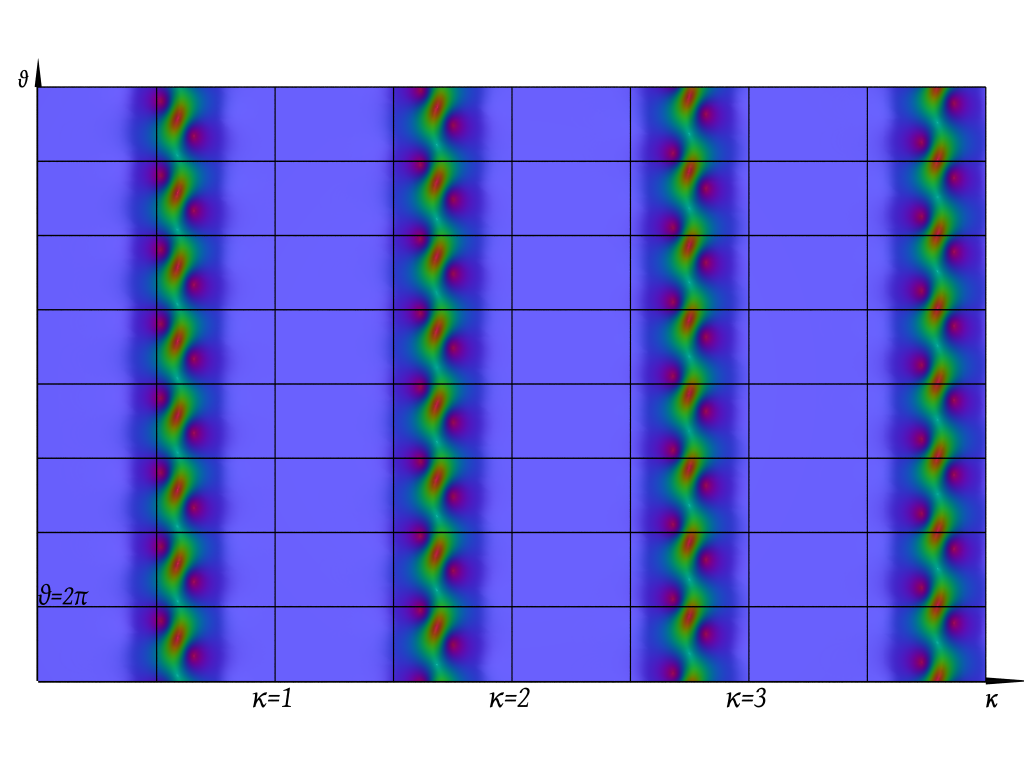}
\end{center}
\caption{The relative amplitude of the corrected solution to the genus-0 solution. The plot is in the variables $\theta,\varkappa$ and the scale is not uniform. The parameters $C,\eta,$ etc. were obtained from an explicit genus-zero solution and are: 
$
\alpha = 0.8120825800+ 1.251939783\,i\ ,\ \ \eta = -.5422408496+1.731859687\, i\ , \ \  
C = .2387253889-.4667042683\, i\ ,\ \ \lambda_0 = -2.358599870+2.555370658\,i\ ,\ \ \varepsilon = 10^{-10}\ ,\ \ x_0 = 0.1\ ,\ \ t_0 = 0.1570482549
$.
It is actually necessary to set such a small epsilon because for $\varkappa \in \N$ we have a discontinuity of order $\sqrt{\varepsilon}$, to be expected since this is the order of 
our approximation and at the integers the correction we have is of the exact same order of the error.
Note that since $\varkappa$ measures distances in the scale $\varepsilon |\ln \varepsilon|$ while $\theta$ in the scale $\varepsilon$, the distance between consecutive ``ranges", 
should be much longer than the transversal separation of the peaks. Also the successive ranges are relatively very thin (of order $1/|\ln \varepsilon|$ relative to the longitudinal separation). 
Therefore the typical {\em size} of the hills is $\varepsilon$ in all directions while the {\em separation} in the longitudinal direction is $\varepsilon|\ln \varepsilon|$. 
This picture is consistent with a full--blown genus-two regime, where the solution is quasi-periodic at the scale $\varepsilon$: as we progress into the genus-2 region the 
separation reduces to the natural scale $\varepsilon$. If we were to plot this in the $(x,t)$ plane, 
the only difference would be  a linear change of coordinates and the ``ranges'' would be parallel to the breaking curve. 
Note that the picture is strictly periodic 
in the $\vartheta$ direction but only {\em quasi-periodic} in the $\varkappa$ direction although it is almost not discernible in the picture: in fact the subsequent ranges are actually slightly shifted.} 
\label{3dpic}
\end{figure}

 The main finding of the paper is the fine structure of the ``shock waves'' along the breaking curve
in the $x,t$-plane, clearly visible on Fig. \ref{Cai}. They turn out to be not ``uniform'',
as one might erroneously conclude looking at Fig. \ref{Cai}.
As shown on Fig. \ref{3dpic}, each wave consists of the sequence of ``peaks'' or ``teeth'' interlaced 
by ``holes'' or ``depressions''. It is remarkable that the mathematical description of this phenomenon is quite similar to the description of 
asymptotic  of ``colonization'' of a new interval by zeroes of certain orthogonal polynomials in \cite{Bertola:Lee, Bertola:wj}.  More precisely:

\begin{itemize}
\item  {the genus zero approximate solution (\ref{zeroappr})}, continued into the genus-two region, gets a finite order correction (w.r.t. $\ve$) only for {\bf (positive)  half--integer values of $\varkappa$},
 {which correspond to the rows of ``teeth''};
\item in between {half-}integers,  the correction is of order $\mathcal O \le(\ve ^{\frac 12\le( {\rm dist}(\varkappa, \Z) - 1\ri)} \ri)$;
\item the correction term exhibits {\em fast oscillations} only in the direction parallel to the breaking curve, i.e., {it} depends on ${\rm e}^{i\frac {\vartheta(x,t)}\ve}$ and it is periodic in $\vartheta$; 
\item all the remaining dependence of the correction is {\em slow}, namely, it depends on $(x,t)$ at a $\mathcal O(1)$ scale;
\item the formula for the correction term is valid with an error term of order $\mathcal O(\sqrt \ve)$;
\item the only dependence on $\ve$ in the correction terms is through the expressions ${\rm e}^{
\pm i\frac {\vartheta(x,t)}\ve}$ and $\ve^{\frac 1 2 \le(1 - K\ri)}$, where $K = \max(0, \lfloor \varkappa \rfloor)$   with $\lfloor \varkappa \rfloor$ denoting the integer part of $\varkappa$. 
\end{itemize}

The picture getting painted is that the first oscillation is of {\em finite amplitude} and the oscillations form ``mountainous ranges'' with flatlands in between of order $\ve\ln \ve$ and peaks separated longitudinally by distances of order $\ve$. Possibly the most descriptive explanation is contained in Fig. \ref{3dpic}, plotted over the $(\varkappa,\vartheta)$ plane which, we remind, is diffeomorphic to a piece of the $(x,t)$ plane around the breaking curve, with $\varkappa$ growing in the transversal direction ($\varkappa$ and $\vartheta$ are not necessarily orthogonal; in fact one could compute their relative angle easily from (\ref{hderiv}))

While the correction is strictly periodic in $\vartheta$, it is {\bf not periodic} in $\varkappa$, despite a superficial inspection of the picture: neither the formula for the correction nor the picture (at a careful inspection) are periodic in $\varkappa$. 

The actual formula  for the correction term is significantly complicated and it is inconvenient to write 
{it} here: it is contained in Thm. \ref{mainthm} and -in quite explicit terms- in Thm. \ref{mainthm2}.
The main point that we observe at this stage is its {\bf universality}, namely, the fact that it does not depend on the details of the initial data. To be more specific, it is an expression that contains only $C, \vartheta, \varkappa, \alpha, \eta$. 

\subsection*{Organization of the paper}
We start with Section \ref{sectreview} where the essentials about the inverse scattering transform are recalled following \cite{TVZ1}. In this section we summarily go through the transformation of the associated Riemann--Hilbert problem along the lines of the Deift--Zhou steepest descent method \cite{DeiftZhou} by introducing the $g$--function (Sect. \ref{sectg}) and the transformation of the RHP (Sect. \ref{modelpr}). 

Section \ref{sectanal} we analyze the differential geometry of the breaking curve and establish the result that $S(x,t)$ is a local diffeomorphism onto the spectral plane (Lemma \ref{Jacobian}). We also analyze the large-time behavior of $S(x,t)$  along such breaking curve (Sect. \ref{longanal}). 

In Section \ref{sectfirst} we start the detailed construction of the approximation to the RHP: the description is made in two steps, mainly for pedagogical reasons. 
In the first step we provide a leading order approximation to the problem and show that the error term fails  necessarily to tend to zero as $\ve\to 0$ for positive half--integer values of $\varkappa$ (eq. \ref{errorestimate2}). This prompts the necessity of a higher order approximation scheme, contained in Sect. \ref{sectimpro}.

The analysis is done firs on the right branch of the breaking curve and hence should be repeated for the left branch; this is done in Sect. \ref{sectleft}. The finding is that the situation is different only in some setup, but the essence of the formulas and of the phenomenon is identical.

The paper culminates in Sect. \ref{sectcorrect} where we provide the final formul\ae\ for the leading and subleading terms asymptotic behavior of the NLS solution (Thm. \ref{mainthm}, Thm. \ref{mainthm2}). We also provide a short discussion about the large-time behavior (along the breaking curve) of the solution.

\section{A short review of the  {zero dispersion limit of the} inverse scattering transform}
\label{sectreview}
%

Given an initial data 
(potential)  $q(x,0)$ for the (\ref{FNLS}) that is  decaying as $x\ra\pm\infty$, the direct scattering transform for the NLS  (\cite{ZS})
produces {the scattering data, namely: the reflection coefficient $r_0(z,\ve)$ and the  points  of discrete
spectrum together with their norming constants (solitons). The time evolution of the scattering data is simple
and well-known. Thus, to find the evolution of a given potential at some time $t$, 
one needs to solve the inverse scattering 
problem at the time $t$. Equivalently, we can stipulate that the initial data is assigned directly 
through the  scattering data;
thus,  one can produce a solution to the NLS (\ref{FNLS}) by choosing  some 
scattering data and solving the inverse scattering problem for $t=0$ (initial data) and for $t>0$ (evolution
of the initial data). The latter approach allows one to avoid solving the direct scattering problem and addressing
many delicate issues associated with it (see \cite{Abel}  for more details). Since we are interested
in studying the generic structure of solutions of the NLS in the transition regions, 
it will be convenient for us to define
a solution to the NLS (\ref{FNLS}) by its scattering, not by its initial, data.}

 In  considering the semiclassical  limit of (\ref{FNLS}), one has to consider the semiclassical limit of the 
corresponding scattering transform. For the case of decaying potentials of type (\ref{IDe}), this limit was discussed in \cite{TVZ3} (inverse scattering) and \cite{Abel} 
(direct scattering), where it was shown to be
a correspondence between 
\be\label{ptlim}
\a(x,t)= -\hf \Phi_x(x)+iA(x,t)
\ee
 on the potential side and 
\be\label{sctlim}
f(z)=f_0(z)-xz-2tz^2
\ee
on the scattering side.
Here $t\geq 0$ is fixed and $f_0(z)$ has the meaning of the ``scaled'' logarithm
of the reflection coefficient $r_0(z,\ve)$ that corresponds to the initial data (\ref{IDe}), that is, 
\be \label{f0}
f_0(z)=\frac{i}{2}\lim_{\ve\ra 0}\ve r_0(z,\ve)~.
\ee
 For example, the limiting $f_0(z)$ that corresponds to the  potential $\cosh (x)e^{\frac{2i}{\ve}\tanh x+i\pi}$, 
is given by (see \cite{TVZ1})
\be\label{f0m=2}
f_0(z)=(1-z)\left[i\frac \pi 2 +\ln(1-z)\right]
+z\ln z+\ln 2+\frac \pi 2 \varepsilon, \ \ \ \ \ \ \mbox{when} \ \Im z\ge 0.  
\ee
Moreover, if $x(\a)$ denotes the inverse function to $\a(x,t)$, at $t$  fixed, then $x(\a)$
and $f_0(z)$ are connected through some  Abel type linear integral transform, see details in \cite{Abel}.

Since we have taken the perspective that we are starting with the scattering data of the form $r_0(z,\epsilon) := {\rm e}^{-\frac{2i}\varepsilon f_0(z)}$ 
we  shall assume that (\cite{TVZ3}):
\begin{itemize}
\item  $f_0(z)$ is analytic in the upper half-plane and has a continuous limit on $\R$;
\item $f_0(z) = \mathcal O(z)$ as $z\to \infty$; 
\item  there exists an interval  $(\m_-,\m_+)\subset \R$ such that  $\Im f_0(z)>0$ for $z\in (\m_-,\m_+)$ and  $\Im f_0(z)<0$ for $z\in (-\infty,\m_-)\cup(\m_+,\infty)$;
\item  $\Im f_0(z),~z\in\R,$ has simple zeroes at $\m_\pm$ and its values are separated from zero outside
neighborhoods of $\m_\pm$.
\end{itemize}

 Due to the Schwarz-symmetry of the scattering data for the NLS, $f_0(z)$ is also a Schwarz-symmetrical
function, i.e., $f_0(\bar z)=\ov{f_0(z)}$; generically, $\Im f_0(z)$ has a jump across the real axis.
 The above assumptions are not overly restrictive: it was shown that  in the case of a  solitonless potential the formula  
(\ref{f0}) yields such an $f_0(z)$   after some process of analytic extension, or ``folding'' of certain logarithmic branch-cuts onto the real axis.
In fact, for our goals it is sufficient to replace the analyticity requirement by a weaker condition:
the contours of the RHP for the $g$-function (which will be introduced in the next section) always lie within the domain of analyticity of $f_0(z)$ in $\C_+$.
Generically, $r_0(z)$ has a {jump discontinuity }  along the real axis due to the discontinuity in $\Im f_0$.

At any time $t$, the inverse scattering problem for  (\ref{FNLS})  (in the solitonless case) with
a fixed (not infinitesimal) $\ve$ is reducible to the following matrix RHP.

\begin{problem}
Find a  matrix $\Gamma(z)$ analytic in $\C\setminus \R$ such that 
\bea\label{rhpgam}
\Gamma_+(z) = \Gamma_-(z) \begin{bmatrix} |r_0|^2 +1  & \ov r_0{\rm e}^{-\frac {2i} \varepsilon \le(2 t z^2 + x z\ri)}\\ r_0 {\rm e}^{\frac {2i}\varepsilon \le(2 t z^2 + x z\ri)} & 1\end{bmatrix} \ ,\ \ z\in \R,\\
\Gamma(z) = \1 +\frac 1  z  \Gamma_1  +\mathcal O(z^{-2})\ ,\ \ \ \ z\to \infty.
\label{gaminf}
\eea
(In the case  with solitons, there are additional jumps across small circles surrounding
the points of discrete spectrum.)  Then 
\be\label{nlssolfull}
q(x,t,\ve):= -2\le(\Gamma_1\ri)_{12}
\ee
is the solution of the initial value problem (\ref{FNLS}) for the NLS equation. 
\end{problem}

The jump matrix for the RHP admits the factorization
\bea\label{factoriz}
\begin{bmatrix} |r|^2 +1  & \ov r\\ r& 1\end{bmatrix} = \begin{bmatrix}  1 & \ov r \\ 0 & 1\end {bmatrix} 
\begin{bmatrix} 1 &0 \\
r&1 \end {bmatrix},
\eea
where $r = r(z;x,t)=r_0(z){\rm e}^{\frac {2i}\varepsilon \le(2 t z^2 + x z\ri)}$.

Inspection of the RHP shows that the matrix $\Gamma(\ov z)^*$ (where $^\star$ stands for the complex--conjugated, transposed matrix)
solves the same RHP with the jump matrix $M(z)$ replaced by $M^{-1}(z)$ and hence 

\bp
\label{propsymmetry}
The solution $\Gamma(z)$ of the RHP for NLS has the symmetry
\be
\Gamma(z)( \Gamma(\ov z))^* \equiv \1~.
\ee
\ep

 In order to study the dispersionless limit $\varepsilon\to 0$, the RHP (\ref{rhpgam})-(\ref{gaminf}) undergoes a 
sequence of transformations (that are briefly recalled in Section \ref{modelpr}) along 
the lines of the nonlinear steepest descent method \cite{DKMVZ, TVZ1}, which reduce it to an RHP that allows for an approximation by the so-called model RHP.
The latter RHP has  piece-wise
constant jump matrices (parametrically dependent on $x,t$)  and, in general,  can be solved explicitly
in terms of the Riemann Theta functions,  
or, in simple cases, in terms of algebraic functions. The $g$-function, defined below,
is the key element of  such a reduction.
\subsection{The $g$-function}
\label{sectg}
Given $f_0(z)$, we introduce  the $g$-function $g(z)=g(z;x,t)$ as the solution to
the following scalar RHP:
\begin{enumerate}
\item  $g(z)$ is
analytic (in $z$) in $\bar\C\setminus 
\g_m$ (including analyticity at $\infty$); 
\item $g(z)$ satisfies the jump condition
\be\label{rhpg}
g_+ + g_-=f_0-xz-2tz^2~~~~ {\rm on}~~\g_{m},
\ee
for  $x\in\R$ and $t\ge 0$, and; 
\item
$g(z)$ has the endpoint behavior
\be\label{modeq}
g(z)=O(z-\a)^{3\over 2}~ + ~{\rm an~analytic~ function~ in~ a~ vicinity~ of~} \a. 
\ee
\end{enumerate}
Here:
\begin{itemize}
\item  $\g_m$ is a bounded Schwarz-symmetrical contour
(called the main arc) with the endpoints $\bar\a, \a$, oriented from $\bar\a$ to $ \a$ and
intersecting $\R$ only at  $\m_+$;
\item  $g_\pm$ denote the values of $g$ on the positive
(left) and negative (right) sides of $\g_m$;
\item the 
function $f_0=f_0(z)$, representing the initial scattering
data, is   Schwarz-symmetrical and H\"{o}lder-continuous on $\g_m$.
\end{itemize}
Taking into the account Schwarz symmetry, it is clear
that  behavior of $g(z)$ at both endpoints $\a$ and  $\bar \a$ should be the same.

Assuming $f_0$ and $\g_m$
are known, {the} solution $g$ to the {scalar} RHP (\ref{rhpg}) without the endpoint condition (\ref{modeq}) 
can be obtained by the Plemelj formula
\be\label{gform}
g(z)={{R(z)}\over{2\pi i}} \int_{\g_m}{{f(\z)}\over{(\z-z)R(\z)_+}}d\z~,
\ee
where $R(z)=\sqrt{(z-\a)(z-\bar\a)}$. 
{ We fix the branch of $R$ by requiring that $\lim_{z\ra\infty}\frac{R(z)}{z}=1$.} If $f_0(z)$ is analytic in some region $\Sscr$ that 
contains $\g_m\setminus\{\m_+\}$, the formula for $g(z)$   can be rewritten as
\be\label{gforman}
g(z)={{R(z)}\over{4\pi i}} \int_{\gt_m}{{f(\z)}\over{(\z-z)R(\z)_+}}d\z~,
\ee
where $\gt_m\subset\Sscr$ is a negatively oriented loop
around $\g_m$ (which is ``pinched'' to $\g_m$ in $\m_+$, where $f$ is not analytic)
that does not contain $z$. 
Introducing function 
$h=2g-f$
we obtain 
\be\label{hform}
h(z)={{R(z)}\over{2\pi i}} \int_{\gt_m}{{f(\z)}\over{(\z-z)R(\z)_+}}d\z~,
\ee
where $z$ is inside the loop $\gt_m$. 
The endpoint condition
(\ref{modeq})  can now be written as
\be\label{modeqh}
h(z)=O(z-\a)^{3\over 2}~~{\rm as}~~z\ra \a,
\ee 
or, equivalently,
\be\label{modeqint}
\int_{\gt_m}{{f(\z)}\over{(\z-\a)R(\z)_+}}d\z=0~.
\ee
The latter equation is known as a {\it modulation equation}. The function $h$ plays a prominent role 
in this paper. Using the fact that the Cauchy operator for the RHP (\ref{rhpg}) commutes with
differentiation, we have 
\be\label{h'form}
h'(z)={{R(z)}\over{2\pi i}} \int_{\gt_m}{{f'(\z)}\over{(\z-z)R(\z)_+}}d\z~,
\ee
where $z$ is inside the loop $\gt_m$. 

 In order to reduce the RHP (\ref{rhpgam})-(\ref{gaminf}) to the RHP with 
piece-wise jump matrices, called the {\em model RHP}, the signs of $\Im h(z)$ in the upper half-plane should satisfy the following conditions:
\bi \label{ineqh}
\item $\Im h(z)$ is negative on both sides of the contour $\g_m$;
\item there exists a continuous contour $\g_c$ (complementary arc) in $\C_+$ that connects $\a$ and $\m_-$, so that 
$\Im h(z)$ is positive along $\g_c$. Since $\Im h(z)>0$ on the interval $(-\infty,\m_-)$,  the point $\m_-$ in
$\g_c$ can be replaced  by any other point of this interval, or by $-\infty$.
\ei

{ Note that the first sign requirement, together  with (\ref{rhpg}), imply that $\Im h(z)=0$
along $\g_m$. Since the signs of $\Im h(z)$ play an important role in the following discussion,
we call by ``sea'' and ``land'' the regions in $\C_+$, where $\Im h(z)$ is negative and positive
respectively. In this language, the complementary arc $\g_c$ goes on ``land'', whereas the main
arc $\g_m$ is a ``bridge'' or a ``dam'', surrounded by the sea, see Fig. \ref{YRHP}.}

\subsection{Reduction to the  model RHP }
\label{modelpr}

{ We start the transformation of the RHP (\ref{rhpgam})-(\ref{gaminf}) by deforming (preserving the orientation) 
the interval $(-\infty,\m_+)$ , which is a part of its jump contour, into some contour $\g^+$ in 
the upper half-plane $\C_+$, such  that $\m_+\in\g^+$. Let $\g_-$ be the Schwarz symmetrical image of $\g_+$. {Using the factorization (\ref{factoriz})} the   
RHP (\ref{rhpgam})-(\ref{gaminf}) {can be reduced to and equivalent one where}:
\bi
\item {the} right factor of (\ref{factoriz}) is the jump matrix on $\g_+$;
\item {the} left factor of (\ref{factoriz}) is the jump matrix on $\g_-$;
\item the jump matrix on the remaining part of $\R$ is unchanged.
\ei
It will be convenient for us to change the orientation of $\g_+$, which causes the change of sign in the off-diagonal entry of the corresponding jump matrix.
{On the interval  $(\m_+,\infty)$ we have  $\Im f_0(z)<0$ and it appears that the jump is exponentially close to the identity jump and hence it 
 is possible to prove that it has no bearing on the leading order term 
of the solution (\ref{nlssolfull}) (as $\ve \ra 0$: see   \cite{TVZ1} for the case when $f_0$ is given by (\ref{f0m=2})
and  \cite{TVZ3} for the general case)}. {Therefore}  the leading order contribution in  (\ref{nlssolfull}) comes from the contour $\g=\g_+\cup\g_-$. 
In the genus zero case, the contour $\g$ contains points $\a,\bar\a$, which divide it into the main arc $\g_m$ (contained between $\bar\a$ and $\a$, and the complementary
arc $\g_c=\g\setminus \g_m$. 
According to the sign requirements (\ref{ineqh}), the contour $\g_m$ is uniquely determined as an arc of the level curve $\Im h(z)=0$ (bridge) that connects
$\m_+$ and $\a$, whereas $\g_c$ can be deformed arbitrarily ``on the land''.
Because of the  Schwarz symmetry
\ref{propsymmetry}, it is sufficient to consider $\g$ only in the upper half-plane, i.e., it is sufficient to consider $\g_+$. 
}

Having found
{the branch-point  $\alpha$, the $g$-function $g(z)$ and
the contour $\gamma_m$,
we introduce} additional contours customarily called ``lenses'' that join $\alpha$ to $\mu_+$ on both sides of $\gamma_m$ 
(and symmetrically down under). These lenses are to be chosen rather freely with the only condition that $\Im h$ must be negative along them (positive in $\C_-$). 
This condition is guaranteed by (\ref{ineqh}).

\begin{figure}[htbp]
\resizebox{0.94\textwidth}{!}{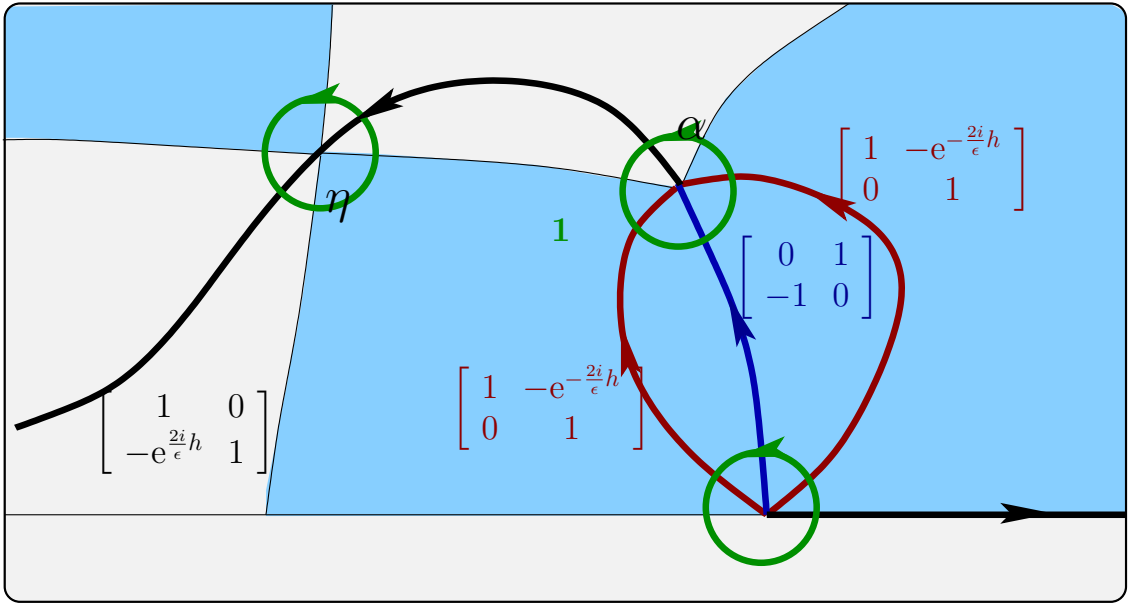}
\caption{The jumps for the  RHP for $Y$. The shaded region is where $\Im h<0$ (the ``sea''). The blue contour is the main arc $\g_m$, the black contour
in $\C_+$ is the complementary arc $\g_c$ and the red contours are the lenses.}
\label{YRHP}
\end{figure}

The two spindle-shaped regions between $\gamma_m$ and the lenses are usually called upper/lower {\bf lips} (relative to the orientation of $\gamma_m$. 
At this point one introduces the auxiliary matrix-valued function $Y(z)$ as follows
\be
Y(z) ={\rm e}^{-\frac{2i}\varepsilon g(\infty) \sigma_3}\Gamma(z)\le\{
\begin{array}{cc} 
\ds {\rm e}^{\frac {2i}\varepsilon g(z) \sigma_3} & \mbox{ outside the lips,}\\
\ds{\rm e}^{\frac {2i}\varepsilon g(z) \sigma_3} \left[
\begin{array}{cc}
1 & -{\rm e}^{-\frac {2i}\ve h(z)}\\
0& 1
\end{array}
\right] & \mbox{ in the upper lip in $\C_+$,}\\
\ds {\rm e}^{\frac {2i}\varepsilon g(z) \sigma_3}\left[
\begin{array}{cc}
1 & {\rm e}^{-\frac {2i}\ve h(z)}\\
0& 1
\end{array}
\right] & \mbox{ in the lower lip in $\C_+$.}
\end{array}
\ri.
\label{GtoY}
\ee
The definition of $Y(z)$ in $\C_-$ is done respecting the symmetry in Prop. \ref{propsymmetry}, namely
\be
Y(z) = (Y(\ov z)^\star)^{-1},\ \ z\in \C_-.
\ee
The jumps for the matrix $Y(z)$ are reported in Fig. \ref{YRHP}.

\paragraph{The model RHP.}
  In the limit $\ve\ra 0$, according to the signs (\ref{ineqh}), the jump matrices on the complementary arc $\g_c$ and on the lenses are approaching the identity matrix $\1$
exponentially fast. Removing these contours from the RHP for $Y(z)$, we  will have only one remaining contour $\g_m$ with the constant jump matrix 
$\left[
\begin{array}{cc}
0 & 1\\
-1& 0
\end{array}
\right]  $ on it. This is the model RHP. Calculating the (1,2) entry of the residue at infinity (see (\ref{nlssolfull})) of the solution to the model RHP, one obtains the leading 
order term of the genus zero solution as follows (\cite{TVZ1})
\be\label{lotg0}
q_0(x,t,\ve)=\Im \a(x,t)e^{-\frac{2i}{\ve}\int_0^x \Re \a_0(s,t)ds}~; 
\ee

To justify removing contours with exponentially small jump matrices, 
one has to calculate the error estimates coming from neighborhoods of points $\a$, $\bar \a$ and $\m_+$ (for $\a$ and $\m_+$, these neighborhoods are shown as green
circles in Fig. \ref{YRHP}). This is often done through the {\em local parametrices}.
 We shall consider the construction of the parametrices near these points as already done and known to the reader. We refer to the literature \cite{TVZ1,DKMVZ}. The only information that we need is that these parametrices allow to approximate the exact solution to within an error term $\mathcal E(z) = \1 + \mathcal O(\epsilon)$.
\section{The analysis near the right breaking curve for NLS}
\label{sectanal}
\begin{figure}[h]
\resizebox{1\textwidth}{!}{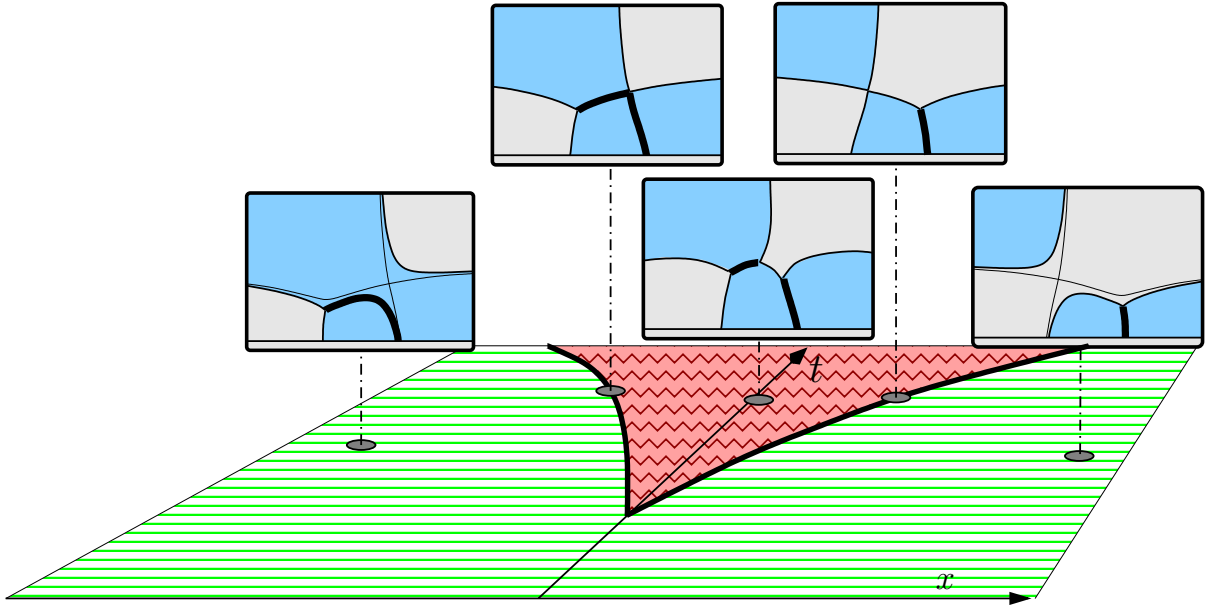}
\caption{The typical zero-dispersion phase diagram for a one-hump initial data. Representative at different points in the $(x,t)$--plane are the level-curves of $\Im (h)$. 
The genus-zero {region} in the $(x,t)$--plane is the lighter area, while the wedge--shaped area is the genus $2$ region. Only the upper-half spectral plane is depicted. The shape of the level 
curves was obtained numerically on a simple example.} 
\label{phasediagram}
\end{figure}

When $x,t$ is on the  breaking curve the asymptotic solution to NLS is about to change {its genus. Although the phenomenon described in this paper is typical  for any
increase of genus, we will focus on the first break, i.e., on the transition} from genus  from  $0$ to genus $2$. This happens by either: 
\begin{itemize}
\item creating a {\bf new band} (right breaking curve in Fig. \ref{phasediagram});
\item splitting an existing band (left breaking curve in Fig. \ref{phasediagram}).
\end{itemize}

Let $(x_0,t_0)$ be on the (smooth part of the) breaking curve; this implies that there is $\eta\in \C_+$
\be
\Im h (\eta, x_0, t_0) = 0\ ,\ \  h' (\eta(x_0,t_0), x_0, t_0) =0\ ,\ \  C= C(x_0,t_0) := -i h'' (\eta(x_0,t_0), x_0, t_0) \neq 0,
\ee
where prime denotes derivative with respect to the spectral parameter $z$; 
moreover, $\Im h(z, x_0, t_0)$ has the required signs (\ref{ineqh}) on both sides of main (bands) 
and complementary (gaps) arcs. 
We keep notation $h(z;x,t)$ for extension  of $h(z;x_0,t_0)$ in a small neighborhood of $(x_0,t_0)$ even if
{\bf  the sign-distribution requirements (\ref{ineqh})  are not satisfied}.
\bd
We denote the critical value 
\be
S(x,t):= h(\eta(x,t), x,t),
\ee
which satisfies $\Im S(x_0,t_0) = 0$. 
\ed

We will now show that $S(x,t)$ is locally a diffeomorphism from the $(x,t)$ plane  near the smooth part of the breaking curve, into the complex $S$--plane. From \cite{TVZ1}, formula (4.46), we have 
\be\label{hderiv}
h_x(z) = R(z)\ ,\qquad h_t(z) = 2(z+a)R(z)\ ,\ \ R(z) = \sqrt{(z-\alpha)(z-\ov \alpha)}, 
\ee
where the square root is chosen to have the behavior $R(z)\sim z$ near infinity (this is the opposite branch compared to the one used in \cite{TVZ1}) and analytic off the main arc. 
Since (by definition of $\eta$) $h'(\eta) =0$ we have that the partial and total derivatives of $S(x,t) = h(\eta(x,t),x,t)$ coincide and hence 
\be
S_x = R(\eta) \ ,\qquad S_t  = 2(\eta+\Re (\alpha) )R(\eta)\ .\label{Sgradient}
\ee
We also have
\bl[Lemma 4.5 in \cite{TVZ1}]
For $t\geq 0$ and for any $x\in \R$ 
\begin{itemize}
\item the function $\Im h_x(z) = \Im R(z)$ is strictly positive in the upper half-plane except for the finite  region enclosed between the main arc and the vertical segment $[\alpha,a]$.
\item The inequality $\Im h_t(z) = 2\Im  (z+a)R(z)$ holds for $|z|$ large enough and to the left of the zero level curve of $\Im h_t(z)$ that emanates from $\a$.
It changes sign when crossing this curve or the main arc.
\end{itemize}
\el
In addition to this we will show that

\bl
\label{Jacobian}
The function $S(x,t)$ is a {\em local} diffeomorphism for $x\in \R$, $t\in \R_+$ with values in $\C\sim\R^2$ except at the point of gradient catastrophe.
\el
{\bf Proof.}
The Jacobian of the map $S(x,t)$ is 
\bea\label{jacobian}
J(x,t) = \det \begin{bmatrix}
\Re \, S_x & \Re\,S_t\\
\Im \, S_x & \Im \,S_t
\end{bmatrix} = \Re \, S_x\Im \,S_t -     \Re\,S_t\Im \, S_x = \Im \le[\ov {S_x} S_t\ri]  = 2\Im (\eta+a)|R(\eta)|^2 = \cr
= 2\Im(\eta) |R(\eta)|^2.
\eea
Thus the Jacobian does not vanish on the upper half-plane and $S$ is a diffeomorphism as long as $\eta\neq \alpha$.
\QED

\bt
\label{breakingcurve}
Assume that on the breaking curve $t(x)$ there is a unique gradient-catastrophe  point $(x_0,t_0)$,
i.e., $\eta(x_0,t_0)=\a(x_0,t_0)$. Then,
outside of  any (small) neighborhood of the gradient-catastrophe  point $(x_0,t_0)$,
the function $S(x,t)$ maps a strip-like neighborhood of  the left breaking curve {\em diffeomorphically} to a strip around the negative real $S$--axis with the genus $2$ phase mapped 
to $\Im S>0$, while it maps a corresponding neighborhood of the right breaking curve to a strip around the {\em positive} real $S$--axis, with the genus $2$ phase mapped to $\Im S<0$. 
\et
{\bf Proof.} The full derivative of $S$ along the breaking curve $t(x)$ is
\bea\label{fullx}
\frac{d}{dx}S(x,t)=\frac{d}{dx}h(\eta(x,t),x, t(x)) 
= h'(\eta, x,t)\eta_x+h_x(\eta, x,t)+ h_t(\eta, x,t)t'(x)=\cr=
h_x(\eta, x,t)+ h_t(\eta, x,t)t'(x)\neq 0.
\eea
The vectors  $h_x(z),h_t(z)$ in $\C$ are not parallel to each other for any $z\in \C^+\setminus\{\a\}$.
But $S(x,t)$ is a real-valued function along $t(x)$, so $\frac{d}{dx}\Re S(x,t) \neq 0$.
Since $S(x_0,t_0)=0$ and $S(x,t(x))\ra +\infty$ as $x\ra+\infty$ (see (\ref{Sass}), the right
branch of the breaking curve is mapped onto the positive real axis of $S$. Similarly,
the left  branch of the breaking curve is mapped onto the positive real axis of $S$. The required
sign distribution  of $\Im S(x,t)$ in the genus two region follows from diagrams at Figure \ref{phasediagram}.
\QED
According to Thm. \ref{breakingcurve}, the real $S$-axis (up to a small neighborhood of the origin) is diffeomorphic to the two branches of the first breaking curve.

We will consider mostly the right-breaking curve, leaving the (not too different) analysis of the left-breaking curve to later: the genus $2$ region corresponds to {\em negative values} of $\Im S$.
We want to ``explore'' the genus-two region within a scale $\varepsilon |\ln \varepsilon|$ and hence we introduce below a conveniently normalized {\bf exploration parameter} $\varkappa$.  
\bd[Exploration parameter]
\label{explore}
We define the two functions $\vartheta(x,t)$, $\varkappa(x,t)$ by the relation
\be
2S(x,t) =\vartheta(x,t) + i \varkappa \varepsilon \ln \varepsilon
\label{foliation}
\ee
\ed
This equation defines implicitly a {\em foliation} of the a neighborhood of $(x_0,t_0)$.
The breaking curve is defined by the equation $\varkappa = 0$ and $ \vartheta$ serves as a parametrization of it, thanks to Lemma \ref{Jacobian}. 
The scaling in (\ref{foliation}) means that $\varkappa$ parametrizes an infinitesimal strip-like neighborhood of the breaking curve with a  width of order $\varepsilon \ln \varepsilon$. 




We will write 
\be
2i h(z) = -C (z-\eta)^2 (\1 + \mathcal O(z-\eta))  + 2i S\ ,\qquad \eta =\eta(x,t),\ \ S:= S(x,t),\ C:= C(x,t).\label{zoomcoord}
\ee
The complementary arc $\g_c$ that goes through $\eta$ can be deformed so that --  in a  finite neighborhood
of $\eta$  --
\be
\Re( h(z) - S)  \equiv 0 ~~~~~~~~~~~~~~~{\rm when}~~~~~~z\in\g_c.
\ee
\bd[Scaling local coordinate]
Define 
\be
\zeta:= \sqrt {-\frac {2i}\varepsilon (h(z)-S)}
\ee
where the root is chosen so that the complementary arc $\Re (h-S)\equiv 0$ is mapped to the real $\zeta$--axis with the natural orientation.
\ed

The jump on the gap (complementary arc) that goes through $\eta_0$ is 
\be
Y_+  = Y_-
\le[
\begin{array}{cc}
1 & 0 \\
{-}{\rm e}^{\frac {2i}{\varepsilon} h(z)} & 1
\end{array}
\ri] = Y_-\le[
\begin{array}{cc}
1 & 0 \\
{-}{\rm e}^{-\zeta^2 + \frac {2iS}\varepsilon} & 1
\end{array}
\ri]  =  Y_-\le[
\begin{array}{cc}
1 & 0 \\
{-}\varepsilon^{-\varkappa}{\rm e}^{-\zeta^2 -  \frac {2i\Re(S)} \varepsilon} & 1
\end{array}
\ri]=: Y_-\le[
\begin{array}{cc}
1 & 0 \\
{-}\varepsilon^{-\varkappa}
{\rm e}^{-\zeta^2 + i \frac \vartheta \varepsilon} & 1
\end{array}
\ri]  
\ee
Note that the oscillatory term ${\rm e}^{i\vartheta / \varepsilon}$ is not problematic {because it has constant modulus}, but the term $\varepsilon ^{-\varkappa}$  diverges for $\varkappa>0$.
\subsection{Long time asymptotics { of $S(\eta)$ along the breaking curve}}
\label{longanal}
Let us recall that the {scattering} data $f_0(z)$ is analytic in $\C_+$ (solitonless case)
and  that the corresponding $w(z)=\Im f_0(z){\rm sign}(z-\m_+)$, restricted to
$z\in\R$, is admissible in the sense of \cite{TVZ3}. {We recall that the gist of the notion of admissibility for $w$ is that } $|r_0(z)|$
has exponential growth on the interval $(\m_-,\m_+)$ as $\varepsilon\ra 0$, exponential decay outside 
$[\m_-,\m_+]$, and $w$ is piece-wise continuously differentiable with $w'(\m_\pm)\ne 0$. As it was mentioned above, the condition of {the absence of solitons} is not essential
if the contour $\g$ of the matrix RHP can be chosen so that it does not intersect 
 {singularities of $f_0(z)$.}

Under the above assumptions and using (\ref{h'form}), (\ref{hform}), the system of equations for the point $\eta=\eta(x,t)$ and the breaking curve $t=t_0(x)$ can be written as
\begin{eqnarray}
\label{Gxtaint}
4t+\frac{1}{\pi}\int_{-\infty}^\infty
\frac{w'(\z)d\z}{(\z-\eta)|R(\z)|}&=0\\
\Im \left\lbrace R(\eta)\left[2t(\eta+a)+x+\frac{1}{\pi}\int_{-\infty}^\infty
\frac{w(\z)d\z}{(\z-\eta)|R(\z)|}\right]\right\rbrace  &=0~. 
\end{eqnarray} 
According to \cite{TVZ3}, Sect. 3.4, this system implies that the right branch of the breaking curve
has the asymptotics
\be\label{asst0}
t(x) = - \frac{1}{4\m_-}x +O(1)\ ,\ \ \ \ {x\to +\infty}
\ee
and  {correspondingly}
\begin{equation}\label{asseta}
\eta=\left( \m_- + O(t^{-1}) \right) +
i\left( \frac{w'(\m_-)}{4t}[1+o(1)]\right) \ ,{\qquad t \to \infty}.
\end{equation}
Moreover {it was shown in   Lemma 3.6 \cite{TVZ3}  that} the branch-point $\a(x,t)$  {approaches} $\m_+$ as $x,t\to\infty$
along the right branch of the breaking curve {according to the following asymptotic formula}
\begin{equation} 
\label{bexp}
b:=\Im \a=e^{ \frac{2\pi t (\m_+-\m_-)}{w'(\m_+)}+O(1)}~.
\end{equation}
The two later equations, together with Proposition 4.3 from \cite{TVZ2}, applied to
\be\label{hz}
h(\eta)= R(\eta)\left[2t(\eta+a)+x+\frac{1}{\pi}\int_{-\infty}^\infty
\frac{w(\z)d\z}{(\z-\eta)|R(\z)|}\right]
\ee
yield 
\be\label{Sass}
S(\eta) \sim 4t\m_-(\m_- - \m_+)
\ee
as $\eta\ra\infty$ along the right branch of the breaking curve.
Using
\be\label{h''form}
h''(\eta)={{R(\eta)}\over{2\pi i}} \int_{\gt_m}{{f'(\z)}\over{(\z-\eta)^2R(\z)_+}}d\z~,
\ee
we can, in a similar way, calculate    
\be\label{Cass}
C(x,t) \sim -4it
\ee
along the right branch of the breaking curve. Similar asymptotic analysis can be done for
the left branch of the breaking curve.

In {the case of $f_0(z)$ given by (\ref{f0m=2})}, using the explicit expression for $h''(z)$ given in \cite{TVZ1},  Section 4.5, we obtain
\be\label{Cm=2}
S(\eta)=-2tR(\eta)\left[\eta+a \right] + \sinh^{-1}\frac{\eta-a}{b}-i\frac \pi 2~.
\ee
\section{First approximation}
\label{sectfirst}
\subsection{Outer parametrix}
\label{Outerspace}
The construction of the outer parametrix, namely, the solution of the model problem in Section \ref{modelpr}, is accomplished along the lines of \cite{Bertola:Lee}. 
The outer parametrix (aka {\em model problem}) is the unique solution of the RHP
\bea
\Psi_+ (z) &\& = \Psi_-(z) \le[\begin{array}{cc} 0 & 1 \\ -1 & 0\end {array}\ri] \ \ \ on \ \ (\alpha,\ov \alpha)\\
\Psi (z) &\& = \mathcal O((z-\alpha)^{-\frac  1  4})\qquad
\Psi (z)  = \mathcal O((z-\ov \alpha)^{-\frac  1  4})\\
\Psi(z) &\& = \1 + \mathcal O(z^{-1}), \qquad z\to\infty.
\eea
$\Psi(z)$  is explicitly given by (\ref{model0}) below.
Following the ideas in \cite{Bertola:Lee} we introduce the so--called {\bf Schlesinger transformation} of the model problem
\begin{problem}[Schlesinger-chain of model problems]
\label{KRHP}
Find a matrix $\Psi_{_K}(z)$ analytic off $[\alpha,\ov\alpha]\cup\{\eta,\ov\eta\}$,  $\eta\not\in[\alpha,\ov\alpha]$, such that 
\bea
\Psi_{_K+}(z)  &\&= \Psi_{_K-}(z) \le[\begin{array}{cc} 0 & 1 \\ -1 & 0\end {array}\ri] \ \ \ on \ \ (\alpha,\ov \alpha)\\
\Psi_{_K} (z) &\&= \mathcal O((z-\alpha)^{-\frac  1  4})\qquad
\Psi_{_K} (z) = \mathcal O((z-\ov \alpha)^{-\frac  1  4})\\
\Psi_{_K}(z) &\&= \1 + \mathcal O(z^{-1})\ ,\qquad z\to\infty\\
\Psi_{_K}(z) &\& = \mathcal O(1) (z-\eta)^{-K\sigma_3}\ ,\qquad z\to \eta \label{ABK}\\
\Psi_{_K}(z) &\& = \mathcal O(1) (z-\ov \eta)^{K\sigma_3}\ ,\qquad z\to \ov \eta.\label{ABKoveta}
\eea
\end{problem}
We have formulated the problem assuming that there is only one main arc (cut) but the modifications in the case of multiple arcs is straightforward (see \cite{Bertola:Lee}); the main point is that we have modified the model problem {\em only} by adding the behavior (\ref{ABK}) and (\ref{ABKoveta}) near the points where a new cut is about to emerge.

\paragraph{Uniformization of the plane {without} the cut.}
Consider the complex $z$--plane with a slit on the segment $[\alpha,\ov \alpha]$. It is elementary to show that the map 
\be
z(\lambda) = \frac {\alpha - \ov \alpha}{4i} \le(\lambda - \frac 1 {\lambda}\ri) + \frac {\alpha + \ov \alpha}{2}=\frac{b}{2}\le(\lambda - \frac 1 {\lambda}\ri)+a,\label{uniformization}
\ee
\begin{wrapfigure}{r}{0.4\textwidth}
\resizebox{0.4\textwidth}{!}{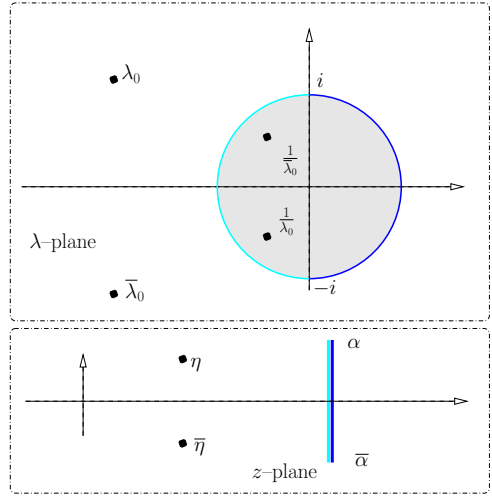}
\caption{The uniformizing plane: the shaded region is the unphysical sheet.}
\end{wrapfigure}
where $\a=a+ib$,
maps {univalently}  and biholomorphically the outside  of the unit disk in the $\lambda$-plane $\{|\lambda| > 1\}$ to $\C\setminus [\alpha,\ov \alpha]$, with the unit $\lambda$-circle being mapped onto the segment two-to-one. The points $\lambda= \pm i$ are mapped to $\alpha, \ov \alpha$ respectively. If we choose another (simple) curve joining $\alpha$ to $\ov \alpha$ then its counter-image in the $\lambda$-plane will be some deformed curve passing through $\lambda = \pm i$ and encircling the $\lambda$--origin. 
Note the symmetry
\be
z(-\lambda^{-1}) = z(\lambda).
\ee
Therefore the outside and inside of $|\lambda|=1$ are conformally mapped to two copies of $z\in \C\setminus [\alpha,\ov \alpha]$.
We will refer to the region $|\lambda|>1$ as the {\bf physical sheet}, as opposed to the {\bf unphysical sheet} $|\lambda|<1$.

The points $z=\eta,\ov \eta$ have  unique pre-images $\lambda_0,\ov \lambda_0$ on the physical sheet
\be
z(\lambda_0) = \eta\ ,\qquad z(\ov \lambda_0) = \ov \eta\ ,\qquad |\lambda_0|>1.\label{lambda0}
\ee
We then have
\bp
\label{propPsiK}
The solution of the RHP (\ref{KRHP}) exists (unique) for any  $K\in \Z$ and it is given by 
\be
\Psi_{_K} =\le( \frac{\ov \lambda_0}{\lambda_0} \ri)^{K\sigma_3}\Psi_{_0} (z)
\le(\frac { \lambda - \lambda_0}{\lambda\lambda_0 + 1}\ri)^{-K\sigma_3}
\le(\frac { \lambda - \ov \lambda_0}{\lambda\ov\lambda_0 + 1}\ri)^{K\sigma_3}\Big|_{\lambda =\lambda(z)}~,
\label{Psik}
\ee
where $\lambda(z)$ is the determination on the physical sheet of $z\in \C\setminus [\alpha, \ov \alpha]$ and $\lambda_0$ (defined in (\ref{lambda0}))  is the counter-image of $\eta$ on the physical sheet.
The matrix $\Psi_{_0}$ is given by 
\be\label{model0}
\Psi_{_0}(z):= \frac 1 2 
\begin{bmatrix}
-i & -1\\
1 & i 
\end{bmatrix}
\le( \frac {z-\alpha}{z-\ov \alpha} \ri)^{\sigma_3/4}
\begin{bmatrix}
i & 1\\  -1 & -i
\end{bmatrix}=\le( \frac {z-\alpha}{z-\ov \alpha} \ri)^{\sigma_2/4}~.
\ee
\ep
{\bf Proof.}
It is well known that the matrix $\Psi_{_0}$ solves the RHP for $K=0$. 
Since the interchange of sheets corresponds to the map $\lambda\mapsto -\frac 1 \lambda$ one immediately verifies that the Szeg\"o\ function 
\be
D(z) := \frac{\lambda\lambda_0 + 1} { \lambda - \lambda_0}
\frac { \lambda - \ov \lambda_0}{\lambda\ov\lambda_0 + 1}
\ee
satisfies $D_+(z)D_-(z) =1$ on the cut, and $\lim_{z\to \infty}D(z) = \frac {\lambda_0}{\ov \lambda_0}$. This is the reason for the left normalization in eq. \ref{Psik}.
\QED

In terms of $\l=\l(z)$, the matrix $\Psi_{_0}$ is given by
\be\label{Psi0l}
\Psi_{_0}(z)=\frac{1}{\sqrt{\l^2+1}}\left( \l\1-i\s_2\right) =\frac{1}{\sqrt{\l^2+1}}
\begin{bmatrix}
\l & -1\\
1 & \l 
\end{bmatrix}
\ee

\begin{remark}
Since $\frac{z-a}{b}=\le(\lambda - \frac 1 {\lambda}\ri)$, the uniformizing variable $\l$ is related 
with the hyperbolic variable $v$, defined by $\sinh v=\frac{z-a}{b}$ from Sect. 4.5, \cite{TVZ1},through
$\l=e^v$. Thus,
\be\label{lambexp}
\l(z)=e^v=\frac{z-a+R(z)}{b},
\ee
where the square root $R$ is chosen so that $\l(z)$ is on the physical sheet.
\end{remark}

\bl
The matrices $\Psi_{_K}$ are related by 
\bea
\Psi_{_{K+1}}(z) = R_{_K}(z)\Psi_{_K}(z)\ ,\ \ \ R_{_K}(z) = \1 - \frac {C_K}{z-\eta}  - \frac {\wh C_K}{z-\ov \eta}\, , \label{Schles}\qquad 
R_K^{-1} =  \1 - \frac {Q_K}{z-\eta}  - \frac {\wh Q_K}{z-\ov \eta}
\eea
\el
{\bf Proof.}
The matrices $\Psi_K$ all have the same jumps and the same asymptotic behavior at the branch-points $\alpha,\ov \alpha$ and at infinity.  Thus $R_K:= \Psi_{_{K+1}} \Psi_{_K}^{-1}$ is a holomorphic matrix (without jumps) with possibly isolated singularities only at $\eta ,\ov \eta$. The asymptotic behaviours of $\Psi_{_{K+1}}$ and $\Psi_{_K}$ at these two points imply that $R_K$ may have at most simple poles there. The same reasoning can be applied to $\Psi_{_K} \Psi_{_{K+1}}^{-1}$, whence the formula for $R^{-1}_K$. \QED

It is possible to give explicit formul\ae\ for the matrices $C_K, \wh C_K, Q_K, \wh Q_K$ in terms of the columns of $\Psi_K$ but they are of no immediate interest.

\subsubsection{Reality}
The matrices $\Psi_K(z)$ can be written as 
\be
\Psi_{_K}(z) = \Phi_K(z)(z-\eta)^{-K\sigma_3}=\bigg[\A_K(z), \B_K(z) \bigg](z-\eta)^{-K\sigma_3} =  
\wh\Phi_K(z)(z-\ov\eta)^{K\sigma_3}=\bigg[\wh \A_K(z), \wh \B_K(z) \bigg](z-\ov \eta)^{K\sigma_3}
\label{ABKs}
\ee
where the columns $\A_K, \B_K$ are analytic at $z=\eta$ and, respectively, the columns $\wh \A_K, \wh \B_K$ are analytic at $z=\ov \eta$.

Note that, since $\det \Psi_K\equiv 1$, we have 
\be
\det\bigg[\A_K(z), \B_K(z) \bigg]   \equiv 1\equiv  \det\bigg[\wh \A_K(z), \wh \B_K(z) \bigg]~.
\ee
Due to the symmetry of Prop. \ref {propsymmetry} we have 
\be\label{symPsi}
\Psi_K(z) (\Psi_K(\ov z))^\dagger\equiv \1~,
\ee
which implies $\Phi_K(z) (\Phi_K(\ov z))^\dagger\equiv 1$ or
\be
\le[\A_K(z) , \B_K(z)\ri]  \equiv \le(\le[\wh \A_K(\ov z), \wh \B_K(\ov z)\ri]^{\dagger}\ri)^{-1}~.
\ee
This is promptly shown to translate into the relations
\be
\A_K(z) \equiv \begin{pmatrix} 0&1\cr-1&0 \end{pmatrix} \ov{\wh \B_K(\ov z)} \ ,\qquad
\B_K(z) \equiv \begin{pmatrix} 0& -1\cr 1&0  \end{pmatrix} \ov{\wh \A_K(\ov z)}~,
\ee
or
\be \label{symPhi}
\Phi_K(z)=\s_2\ov{\wh \Phi_K(\ov z})\s_2,
\ee
which is equivalent to $\Phi_K(z) (\Phi_K(\ov z))^\dagger\equiv \1$.
Similar considerations lead to 
\be \label{symPhi'}
\Phi'_K(z)=\s_2\ov{\wh \Phi'_K(\ov z})\s_2\ ,\qquad \s_2:=  \le[\begin{array}{cc} 0& -i \\ i & 0 \end{array}\ri] 
\ee

We will denote by $\A_K$ the {\em value} $\A_K(\eta)$, and similarly 
\bea
\A_K:= \A_K(\eta)\ ,\ \B_K:= \B_K(\eta)\ , \  \A_K':= \A_K'(z)|_{z= \eta}\ ,\ \B'_K:= \B'_K(z)|_{z= \eta}~,\\
\wh \A_K:= \wh\A_K(\ov\eta)\ ,\ \wh\B_K:= \wh\B_K(\ov\eta)\ , \  \wh\A_K':= \wh\A_K'(z)|_{z=\ov \eta}\ ,\ \wh \B'_K:= \wh \B'_K(z)|_{z=\ov \eta}~.
\eea

This implies the following relations, which will be of use at a later stage in the paper
\bea
\det\le[\A_K', \A_K\ri] = \ov{\det\le[\wh \B'_K,\wh \B_K\ri]}
\qquad
\det\le[\B_K', \B_K\ri] = \ov{\det\le[\wh \A'_K,\wh \A_K\ri]}~, \cr
\det\le[\wh \A_K, \B_K\ri] = \B_K^\dagger \B_K>0\qquad
\det\le[ \A_K, \wh \B_K\ri] = \A_K^\dagger \A_K>0~.
\label{Reality}
\eea
These quantities are computed explicitly in App. \ref{appexplicit}

\subsection{Local parametrices near $\eta,\ov \eta$ and first approximation to the solution}
\label{sectfirstlocal}
We only consider the point $\eta$, with similar considerations being repeated for $\ov \eta$.
The local parametrix must solve the following {\bf mixed RHP} where the adjective ``mixed'' means that the requirements are for both {\bf columns and rows}:
\bea
\mathbf H_{_K+}(z) &\& = \mathbf H_{_K-}(z) \le[
\begin{array}{cc}
1 & 0 \\
{-}
\varepsilon^{-\varkappa}{\rm e}^{i\theta/\varepsilon}{\rm e}^{-\zeta^2} & 1
\end{array}\ri]~;\label {Hjump}\\
\mathbf H_{_K}(z) &\& = (z-\eta)^{K\sigma_3} \mathcal O(1)~;\label{rowsing}\\
\mathbf H_{_K}(z) &\& = \1 + o(1)\ ,\ \ z\in \pa \mathbb D~,\label{boundarydecay}
\eea
where $o(1)$ above means an infinitesimal in $\varepsilon$, uniform w.r.t. $z$ when $z$ belongs to the boundary of the disk. When necessary to specify the values of the parameters $\varkappa, \vartheta$ we will write 
\be
\mathbf H_{_K+}(z) =\mathbf H_{_K+}(z;\varkappa, \vartheta)\ . 
\ee
\subsection{Approximate solution and error analysis}
\label{sectapprox}
Suppose that we are able to solve the mixed RHP in (\ref{boundarydecay}) and define the following matrix
\be
\wt Y(z):= \le\{
\begin{array}{cc}
\Psi_K(z) & \hbox{outside of the local disks}~,\\[10pt]
\Psi_K(z) \mathbf H_K(z)  & \hbox{ within the disks at $\eta,\ov \eta$}~,\\[10pt]
\Psi_K(z)\mathcal A(z) & \hbox{ within the disks at $\alpha, \ov \alpha$}~,
\end{array}
\ri.
\ee
where $\mathcal A(z)$ is the usual Airy-parametrix \cite{DKMVZ}. 
Note that the matrix $\Psi_K(z) \mathbf H_K(z)$ is {\bf bounded} in a neighborhood of $\eta$ because the singular behavior of the columns of $\Psi_K$ (\ref{ABK}) 
is exactly canceled by the singular behavior (\ref{rowsing}) of the rows of $\mathbf H_K$.
In order to evaluate how close the matrix $\wt Y(z)$ is to the matrix $Y(z)$ we take their ratio
\be
\mathcal E(z):= Y(z)\wt Y(z)^{-1}
\ee
and consider the RHP it solves:
\be
\mathcal E(z) \sim \1 + \mathcal O(z^{-1}), ~~~~~~~~~~~~~~~
\mathcal E_+(z)  = \mathcal E_-(z) W(z)~,
\ee
\begin{wrapfigure}{r}{0.2\textwidth}
 \resizebox{0.2\textwidth}{!}{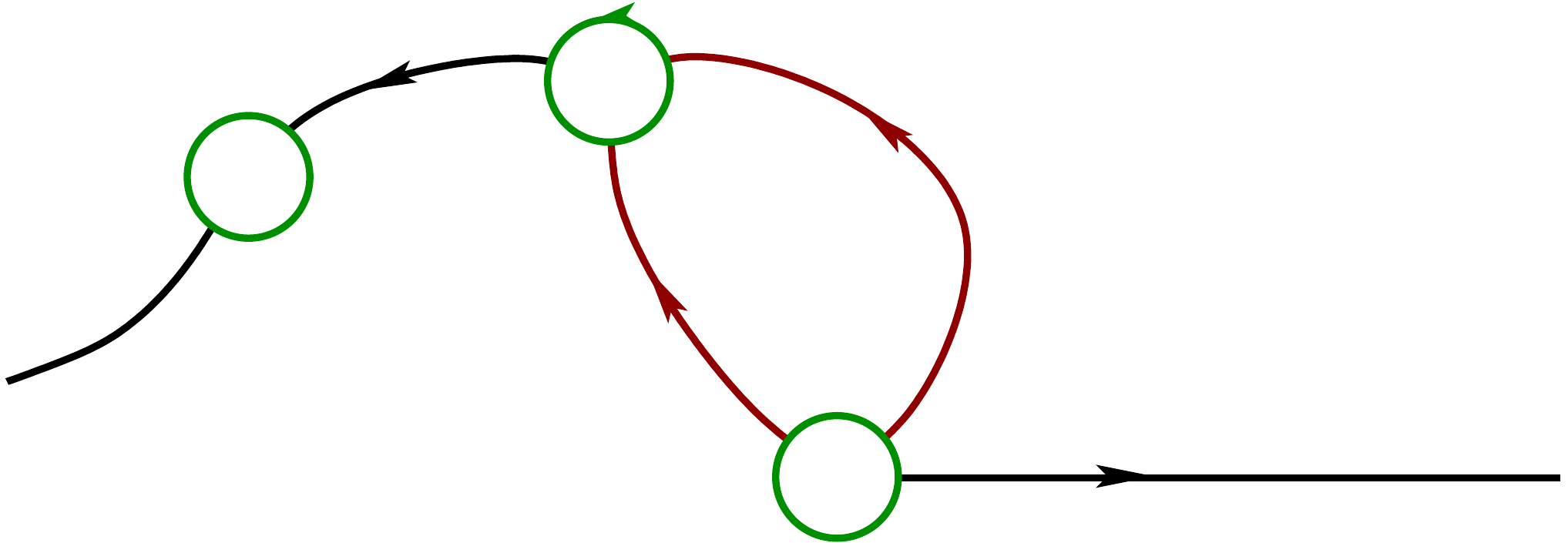}
\caption{The contours for the RHP of the error term (in $\C_+$).}
\label{ERHP}
\end{wrapfigure}

where $W(z)$ is a matrix supported on the contours indicated in Fig. \ref{ERHP}.
In particular, on the boundary of the disk $\mathbb D$ centered at $\eta$ we have 
\be
\mathcal E_+(z) = \mathcal E_-(z) \Psi_K \mathbf H_K(z) \Psi_K^{-1}(z)\ ,\qquad z\in \pa \mathbb D~.
\ee
Since $\Psi_K$ are uniformly bounded for $z\in \pa\mathbb D$ as $\varepsilon\to 0$ (in fact they are independent of $\varepsilon$), 
it follows that the jump of $\mathcal E$ on $\pa \mathbb D$ is $\1 +  o(1)$, where this is the same infinitesimal that appears in eq. (\ref{boundarydecay}) ({and we will promptly see below that this desired infinitesimal is not such when $\varkappa$ is a half--integer, thus requiring a more refined approach}). 
The usual theorem about Riemann--Hilbert problems with small-norm jumps imply that $\mathcal E(z)$ is uniformly close to the identity to within the same infinitesimal. 

We will see in Section \ref{localHk} that it is possible to solve the local problem for $\mathbf H_K$ formulated above with $o(1) = \mathcal O(\varepsilon^{\frac 1 2 - |\varkappa -K|})$  and hence: 
\begin{itemize}
\item the integer $K$ must be chosen as the nearest positive integer to $\varkappa$;
\item it is impossible to solve the problem as formulated for  a half-integer $\varkappa = \frac 1 2 + \mathcal N$, where $\mathcal N\in\N$.
\end{itemize}
We will show below how to solve the problem as formulated, why it is impossible to solve in those exceptional situation and how to circumvent the problem.

\subsection{Solution of the local RHP for $\mathbf H_K$ (\ref {Hjump}).}
\label{localHk}
Let $H_\ell(s)$ be the {\bf monic} Hermite polynomials of degree $\ell$:
\bea
\int_\R H_\ell(s) H_{\ell'}(s) {\rm e}^{-s^2}\d s =\frac  {\ell!  \sqrt{\pi}}{2^\ell} \delta_{\ell\ell'} =: \mu_\ell \delta_{\ell \ell'}\ , 
\qquad H_\ell(s) := (-2)^{-\ell} {\rm e}^{s^2} \frac { \d ^\ell}{\d s^\ell} {\rm e}^{-s^2}~.
\eea
We define the local parametrix as  
\bea
\label{RK}
\mathbf H_K(\zeta)&\&:={\rm e}^{-i\frac \vartheta{2\varepsilon} \sigma_3}\varepsilon^{\frac {\varkappa }  2\sigma_3}\zeta^{K\sigma_3}\le[
\begin{array}{cc}
\ds
\ds\frac {-1}{\mu_{K-1}}  \int_\R \frac{H_{K-1}(s){\rm e}^{-s^{2}}\ds}{s-\zeta}
&
\ds\frac { 
{-}2i\pi}{\mu_{K-1}}H_{K-1}(\zeta) 
\\
\ds 
\frac 1{2i\pi} \int_\R \frac{H_K(s){\rm e}^{-s^{2}}\d s}{s-\zeta}
&
H_K(\zeta) 
\end{array} 
\ri] \varepsilon^{-\frac \varkappa 2\sigma_3} {\rm e}^{i\frac \vartheta {2\varepsilon} \sigma_3}\cr
=&\& 
{\rm e}^{-i\frac \vartheta {2\varepsilon} \sigma_3}
\le[
\begin{array}{cc}
\ds\frac {-\zeta^{K} }{\mu_{K-1}}  \int_\R \frac{H_{K-1}(s){\rm e}^{-s^{2}}\d s}{s-\zeta }
&
\ds \frac {
{-}2i\pi \zeta^K \varepsilon^{\varkappa} }{\mu_{K-1}}H_{K-1}(\zeta) 
\\
\ds 
\frac { \zeta^{-K}\varepsilon^{- \varkappa}} {2i\pi } \int_\R \frac{H_K(s){\rm e}^{-s^{2}}\d s }{s-\zeta}&
\ds \frac{H_K(\zeta)}{\zeta^K}
\end{array} 
\ri]{\rm e}^{i\frac \vartheta {2\varepsilon} \sigma_3}\\
\eea


{\bf Boundedness.} The matrix  $\Psi_K \mathbf H_K$ is bounded in ${\mathbb D}$, because of the  $\zeta^{K\sigma_3}$ factor in (\ref{RK}) cancels out the singularity of $\Psi_K$ (\ref{ABK}).

{\bf Error estimate.} 
Recall that for $z\in \pa \mathbb D$ the zooming coordinate $\zeta$ scales like $\frac 1  {\sqrt{\varepsilon}}$. Therefore 
the matrix $\mathbf H_K$ on the boundary of the disk $\mathbb D$ has the following behavior uniformly w.r.t. $z$
\be
\mathbf H_{K}=\le[
\begin{array}{cc}
1 + \mathcal O(\varepsilon) & \mathcal O  (\varepsilon^{\frac 12 + \varkappa - K})\\
\mathcal O(\varepsilon^{\frac 1 2 - \varkappa + K}) & 1 + \mathcal O(\varepsilon)
\end{array}
\ri]  = \1 + \mathcal O(\varepsilon ^{\frac 1 2 - |\varkappa -K| }),\quad z\in \pa\mathbb D\label{errorestimate2}.
\ee
Note that the diagonal estimates are $1+ \mathcal O(\varepsilon)$ and not just $1+O(\sqrt{\varepsilon})$ due to the fact that the 
Hermite polynomials are even/odd functions according to the parity of their degrees.
For example, for the $(1,2)$ element in (\ref{RK}): $H_{K-1}(\zeta)  = \zeta^{K-1} (1 + \mathcal O(\zeta^{-1}))$  and  on $\pa \mathbb D$ we have $\zeta = \mathcal O(1/\sqrt{\varepsilon})$ and
\be
\frac{2i\pi \zeta^K {\rm e}^{-i\frac \vartheta {\varepsilon}} \varepsilon^\varkappa}{\mu_{K-1}} H_{K-1} = 
\frac {2i\pi \zeta^{2K-1} {\rm e}^{-i\frac \vartheta {\varepsilon}} \varepsilon^\varkappa}{\mu_{K-1}} (1 + \mathcal O(\sqrt{\varepsilon})) = \mathcal O( \varepsilon^{\varkappa - K + \frac 12})~.
\ee
Similarly, for the $(2,1)$ element, using the orthogonality, we have that the Cauchy transform is $\mathcal O(\zeta^{-K-1})$ and hence 
\be
-\frac {\zeta^{-K}  {\rm e}^{i\frac \vartheta {\varepsilon}}\varepsilon^{-\varkappa}}{2i\pi} 
\int \frac {H_K(s){\rm e}^{-s^2}}{s-\zeta} {\rm d} s = \frac {\mu_K}{2i\pi} \zeta^{-2K-1} {\rm e}^{i\frac \vartheta {\varepsilon}} \varepsilon^{-\varkappa} (1+\mathcal O(\zeta^{-1})) = 
\mathcal O(\varepsilon^{K+\frac 1 2 -\varkappa})~.
\ee
We immediately see that --in order to meet the requirement about the decay on the boundary of the disk $\mathbb D$ (\ref{boundarydecay})-- we must have
\bp
The integer $K$ must be the closest integer to $\varkappa$.
\ep

The important observation is that if $\varkappa \in \frac 1 2 + \Z$ then the error term in (\ref{errorestimate2}) does not tend to zero (it is $\mathcal O(1)$).
It is understandable as these values separate regimes where the value of $K$ jumps by one unit and the whole strong asymptotic must changes its form. 
\section{Improved approximation}
\label{sectimpro}
\subsection{Next order improved local parametrix}
\label{sectnextorder}
Inspecting directly and a bit closer  the expression (\ref{RK}) we see that 
\be
\mathbf H_{K}=\le[
\begin{array}{cc}
1 + \mathcal O(\varepsilon) &
\ds \frac {
{-} 2i\pi \zeta ^{2K-1} {\rm e}^{-i\frac \vartheta {\varepsilon}} \varepsilon ^\varkappa}{\mu_{K-1}} + \mathcal O  (\varepsilon^{\frac 32 + \varkappa - K})\\
\frac{
{-}\mu_K}{2i\pi \zeta^{2K +1} {\rm e}^{-i\frac \vartheta {\varepsilon}} \varepsilon^\varkappa}  + \mathcal O(\varepsilon^{\frac 3 2 - \varkappa + K}) & 1 + \mathcal O(\varepsilon)
\end{array}
\ri]  = \1 + \mathcal O(\varepsilon ^{\frac 1 2 - |\varkappa -K| }),\quad z\in \pa\mathbb D\label{errorestimate3}.
\ee
and the terms that are responsible for the loss in the error term (\ref{errorestimate2}) are the off--diagonal ones. In particular:
\begin{itemize}
\item for $K< \varkappa <K+\frac 1 2$ the leading term in the error is the {\bf lower triangular} one;
\item for $K-\frac 1 2 < \varkappa <K$ the leading term in the error is the {\bf upper triangular} one.
\end{itemize}
For this reason we re-define 
\bd
\label{Hk}
The local parametrix $\mathbf H_\varkappa$, $\varkappa \in \R_+$ is defined as 
\be
\mathbf H_\varkappa = \le\{
\begin{array}{cc}
\ds \le[
\begin{array}{cc}
1  &0 \\
\ds \frac{ 
 \mu_K}{2i\pi \zeta^{2K+1}  {\rm e}^{-i\frac \vartheta {\varepsilon}} \varepsilon^\varkappa}  & 1
\end{array}
\ri] \mathbf H_K  & K<\varkappa \leq K + \frac 12
\\[28pt]
\mathbf H_K & \varkappa = K\\[18pt]
\le[
\begin{array}{cc}
1  &\ds 
\frac {2i\pi \zeta ^{2K-1} {\rm e}^{-i\frac \vartheta {\varepsilon}} \varepsilon ^\varkappa}{\mu_{K-1}} \\
0 & 1
\end{array}
\ri]\mathbf H_K & K-\frac 1 2 < \varkappa <K~.
\end{array}
\ri.
\ee
\ed

\paragraph{Parametrix on the disk around $\ov \eta$.} 
Because of the symmetry of Prop. \ref{propsymmetry} we have
\be
\wh {\mathbf H}_\varkappa (\zeta):= \le( \mathbf H_\varkappa(\ov \zeta)^\dagger\ri)^{-1}~.
\ee
In the formul\ae\ below, we will use $\wh \zeta(z):= \ov{\zeta(\ov z)}$.
\subsection{Improved outer parametrix}
\label{sectimprovouter}
Corresponding to the definition of $\mathbf H_\varkappa$ (Def. \ref{Hk}) we need to define the outer parametrix $\Psi_{\varkappa}$ with the requirements 
\bea
&& \Psi_\varkappa(z)_+ = \Psi_{\varkappa}(z)_- \le[
\begin{array}{cc}
0 & 1\\
-1 & 0
\end{array}
\ri]~,
\label{71}\\
&& \Psi_{\varkappa} (z) = \mathcal O((z-\alpha)^{-\frac  1  4})\qquad
\Psi_{\varkappa} (z) = \mathcal O((z-\ov \alpha)^{-\frac  1  4}),\\
&&\Psi_{\varkappa}(z) = \1 + \mathcal O(z^{-1}),\\
&&\Psi_{\varkappa}(z)\mathbf H_{\varkappa} (\zeta) =\mathcal O(1)\ ,\ \ z\in \D\label{bound1},\\
&& \Psi_{\varkappa}(z)\wh {\mathbf H}_{\varkappa} (\wh \zeta) = \mathcal O(1) \ ,\ \ \ z\in \wh \D.\label{bound2}
\eea
In correspondence with Def. \ref{Hk} we will pose the following {\bf Ansatz} for $\Psi_\varkappa$.
\bt
\label{thmpsikappa}
The solution of the Riemann--Hilbert problem  (\ref{71})-(\ref{bound2}) always exists and is given by
\be
\Psi_\varkappa(z) = \le\{
\begin{array}{cc}
\ds T_{K, \varkappa}(z)\Psi_K(z)=\le(\1 - \frac {G_{K,\varkappa}}{z-\eta} - \frac {\wh G_{K,\varkappa}}{z-\ov \eta}\ri)\Psi_K(z)  & K < \varkappa\leq K+\frac 12 \\
\ds \Psi_{_K}(z)  &  \varkappa = K\in \N\\
\ds  U_{K, \varkappa}(z)\Psi_K(z) = \le(\1 - \frac {F_{K,\varkappa}}{z-\eta} - \frac {\wh F_{K,\varkappa}}{z-\ov \eta}\ri)\Psi_K(z)  & K-\frac 12  < \varkappa< K,
\end{array}\ri.
\label{Ansatz}
\ee
where
\bea
G_{K,\varkappa} (\bullet)&\& = \frac {u_K \det [ \bullet, \B_K] }{ \Big|1+ u_K\det[\B_K', \B_K]\Big|^2
   +   \frac {|u_K |^2  \|\B_{_K} \|^4 } {4\Im(\eta)^2} }
\le(\le(1 + \ov u_K \ov{\det[ \B_K' ,  \B_K]}\ri) \B_K - \frac {\ov u_K \|\B_K\|^2}{2\Im(\eta)} \sigma_2\ov  {\B_K} \ri),
\label{Gkappas}
\\
F_{K,\varkappa}(\bullet) &\& = \frac{ \ell_K \det[\A_K,\bullet] } 
{\Big|1 - \ell_K \det[\A'_K,\A_K] \Big|^2  +\frac{ |\ell_K|^2 \|\A_K\|^4}{ 4\Im(\eta)^2} }
\le(\le(
1- \ov \ell_K \ov{\det[\A_K',\A_K]} \ri) \A_K - \frac{ \ov \ell_K\|\A_K\|^2}{2\Im(\eta)}  \s_2 \ov {\A_K}\ri),\label{Fkappas}
\\
&\& \label{Gsym1}
\wh G_{K,\varkappa} =\s_2\ov{G_{K,\varkappa}}\s_2, \qquad \wh F_{K,\varkappa} =\s_2\ov{F_{K,\varkappa}}\s_2,
\\
&\& u_K:=
\frac {K! \sqrt{\pi} C^{-K-\frac 12}}{2^{K+1} i\pi }\varepsilon^{\frac 1 2 -\varkappa + K }  {\rm e}^{i\frac \vartheta {\varepsilon}},\qquad
\ell_K :=  
 \frac {2i\pi  {\rm e}^{-i\frac \vartheta {\varepsilon}} \varepsilon^{\varkappa - K + \frac  1 2}} {\mu_{K-1} C^{-K+\frac 1 2}}. \ .
\label{uk}
\eea
\et
The proof can be found in App. \ref{proofs}.
{\begin{remark}
The idea behind this Ansatz is the following: the outer parametrices $\Psi_K$ and $\Psi_{K+1}$ are related by a finite Schlesinger transformation of the form (\ref{Schles}) and we know that they provide the correct leading behavior for $\varkappa\in (K-\frac 1 2, K+\frac 12) $ and $\varkappa \in (K+\frac 12 , K+\frac 3 2 )$, respectively. The discontinuity in this leading-order asymptotic behavior occurs at $\varkappa = K+\frac 1 2$, where the error term as per (\ref{errorestimate2}) becomes of order $1$; the idea is thus to introduce some sort of interpolation between the two outer parametrices of the same form of a finite Schlesinger transformation, letting the coefficients matrices be dependent on $\varkappa$ in such a way that their order becomes $1$ as $\varkappa$ crosses the critical value.
\end{remark}}

%

Suppose $\kappa> K+\frac 1 2$ in the formula (\ref{Gkappas})  for $G_{K,\varkappa} \ (\wh G_{K,\varkappa})$; then one sees that $u_K$ diverges as $\varepsilon\to 0$. The matrix $G_{K,\varkappa}$ ($\wh G_{K,\varkappa}$), however, has a regular limit $G_{K,\varkappa} \to C_K\ ,\ (\wh G_{K,\varkappa}\to \wh C_K)$, 
while for $\varkappa <K-\frac 1 2$ we have $F_{K,\varkappa}\to Q_K\ ,\ (\wh F_{K,\varkappa} \to \wh Q_K)$
where $C_K, \wh C_K, Q_K, \wh Q_K$ have been introduced in (\ref{Schles}); indeed the defining equations for $G,\wh G_{K,\varkappa}$ turn into those of $C_K, \wh C_K$ (ditto for $F_{K,\varkappa}, \wh F_{K,\varkappa}$ versus $Q_K, \wh Q_K$). This means that $G_{K,\varkappa}, F_{K,\varkappa}$ are some sort of {\bf interpolation} between Schlesinger transformations.
\paragraph{Continuity at $\varkappa =K+\frac 1 2$.}
We must check that the matrix $\Psi_\varkappa$ as defined in eq. (\ref{Ansatz}) is continuous as $\varkappa =K+\frac 12$, at least within the error $\mathcal O(\sqrt{\ve})$. In fact we are going to see below (Prop. \ref{propcont}) that this continuity holds exactly, namely,
\be
\lim_{\varkappa\to K+\frac 1 2 -0} \Psi_{\varkappa} = \lim_{\varkappa\to K+\frac 1 2 +0} \Psi_{\varkappa}
\ee
or, equivalently, 
\be
\le(\1 - \frac {G_{K,\varkappa}}{z-\eta} - \frac {\wh G_{K,\varkappa}}{z-\ov \eta} \ri)_{\varkappa=K+\frac 1 2} \Psi_K(z) = \le(\1 - 
\frac {F_{K+1,\varkappa}}{z-\eta} - \frac {\wh F_{K+1,\varkappa}}{z-\ov \eta} \ri)_{\varkappa=K+\frac 1 2} \Psi_{K+1}(z).\label{727}
\ee
Even more is true as eq. (\ref{727})  holds for {\bf any value} of $\varkappa$. To show this we recall that  $\Psi_{K+1} = R_K \Psi_K$ (as per eq. \ref{Schles}). 
Thus, we prove the following
\bp
\label{propcont}
{\em For any value of $\varkappa\in\R$, } the matrices $G_{K,\varkappa}, \wh G_{K,\varkappa}, F_{K+1,\varkappa}, \wh F_{K+1,\varkappa}$ are related with the $C_K, \wh C_{K+1}$ by
\be
\1 - \frac {G_{K,\varkappa}}{z-\eta} - \frac {\wh G_{K,\varkappa}}{z-\ov \eta} = \le(\1 - \frac {F_{K+1,\varkappa}}{z-\eta} - \frac {\wh F_{K+1,\varkappa}}{z-\ov \eta} \ri) \le(\1 - \frac{C_K}{z-\eta} - \frac {\wh C_K}{z-\ov \eta}\ri).\label{conteq}
\ee
In particular, we have 
\be
G_{K,\varkappa} + \wh G_{K,\varkappa}  = F_{K+1,\varkappa} +\wh F_{K+1,\varkappa} + C_K + \wh C_K\ .
\label{DeltaGK}
\ee
\ep
The proof can be found in Appendix \ref{proofs}.

Thanks to Proposition \ref{propcont} we conclude also that  in Theorem \ref{thmpsikappa}
\be
T_{K,\varkappa}(z) \Psi_{K}(z) \equiv U_{K+1, \varkappa}(z) \Psi_{K+1}(z)\ ,\ \  \ \forall \varkappa,\ z,\ \ve,
\ee
which allows us to dispose of the necessity for the $F_K$ altogether and rewrite  Theorem \ref{thmpsikappa} as 
\bc\label{corpsikappa}
The solution to the RHP (\ref{71})-(\ref{bound2}) given by Theorem \ref{thmpsikappa} can be equivalently written as
\be
\Psi_{\varkappa}(z) = \le(\1 - \frac {G_{K,\varkappa}}{z-\eta} - \frac {\wh G_{K,\varkappa}}{z-\ov \eta}\ri) \Psi_K(z)\ ,\qquad
K = \max(0,\lfloor \varkappa\rfloor)\ ,
\ee
with the matrices being defined in (\ref{Gkappas}).
\ec
We remind the reader that $\Psi_K$ (with integer subscript) had been defined independently in Proposition \ref{propPsiK}.
\br
We also remark that the new definition in Cor. \ref{corpsikappa} is {\bf not continuous} in $\varkappa$ at the integers (but continuous at the half-integer). 
However the discontinuity is to within $\mathcal O(\sqrt{\ve})$. More precisely,
\be
\lim_{\varkappa \to K^-} \Psi_{\varkappa}(z)  =\le(\1 +\mathcal O(\sqrt{\ve}) \le(\frac 1{z-\eta} + \frac 1{z-\ov \eta}\ri)\ri) \lim_{\varkappa \to K^+} \Psi_{\varkappa}(z)
\ee
\er 
with the $\mathcal O(\sqrt{\ve})$ factor being independent of $z$. This discontinuity is of no concern because the correction to the final approximation is within our overall error estimate.
\section{The left breaking curve}
\label{sectleft}
The situation at the left breaking curve (Fig. \ref{phasediagram}) is at first sight quite different because it is the main arc that it is splitting into two, rather than  the complementary arc. However, as we will presently see, the phenomenon is essentially identical: the informal reason is that the choice of main and complementary arcs in the construction of the function $h$ is somewhat conventional, depending on the overall sign of $h$ at infinity and the placement of the cuts of the $g$--function.  The difference is nevertheless sufficiently stark to grant a separate analysis.

\begin{figure}[h]
\centerline{\resizebox{0.8\textwidth}{!}{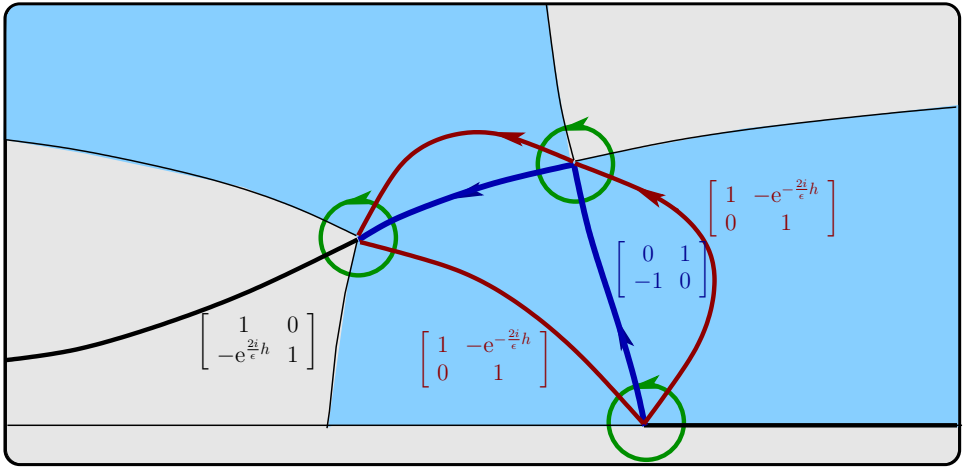}}
\caption{The typical shape of the main, complementary arcs and lens contours at the precise point on the left breaking curve.}
\end{figure}
\begin{wrapfigure}{r}{0.5\textwidth}
\resizebox{0.5\textwidth}{!}{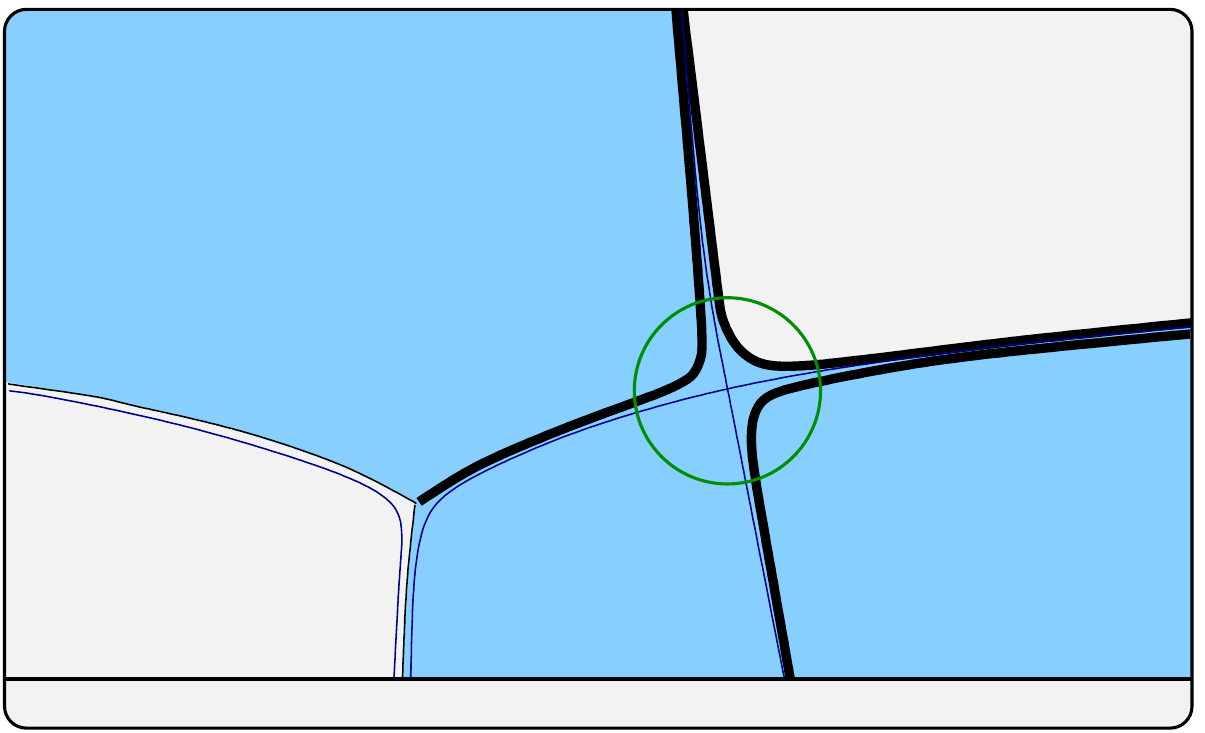}
\caption{The main arc has split into two and we are forced to introduce a new main arc on the zero-level set of $\Im(h)$.}
\label{ViolationBreak}
\end{wrapfigure}

At the exact point of break the function $h$ near the point $\eta$ (where the arc is about to split) can be written 
\be
2ih(z) = \le\{ \begin{array}{cc}
-C (z-\eta)^2(\1 + \mathcal O(z-\eta)), & z\in \D_1,\\
C (z-\eta)^2(\1 + \mathcal O(z-\eta)), & z\in \D_0,
\end{array}
\ri.
\ee
where the region $\D_1$ is the region (within the local disk) to the right of the two main arcs and $\D_0$ its complement. 
If we let evolve the level curves according to the modulation equations while respecting the 
 {branch-cuts} we would have the situation depicted on Fig. \ref{ViolationBreak}, with two pairs of close cuts running off to infinity (i.e. a new main arc appears to compensate the split of the original one).

\begin{figure}[h]
\resizebox{0.5\textwidth}{!}{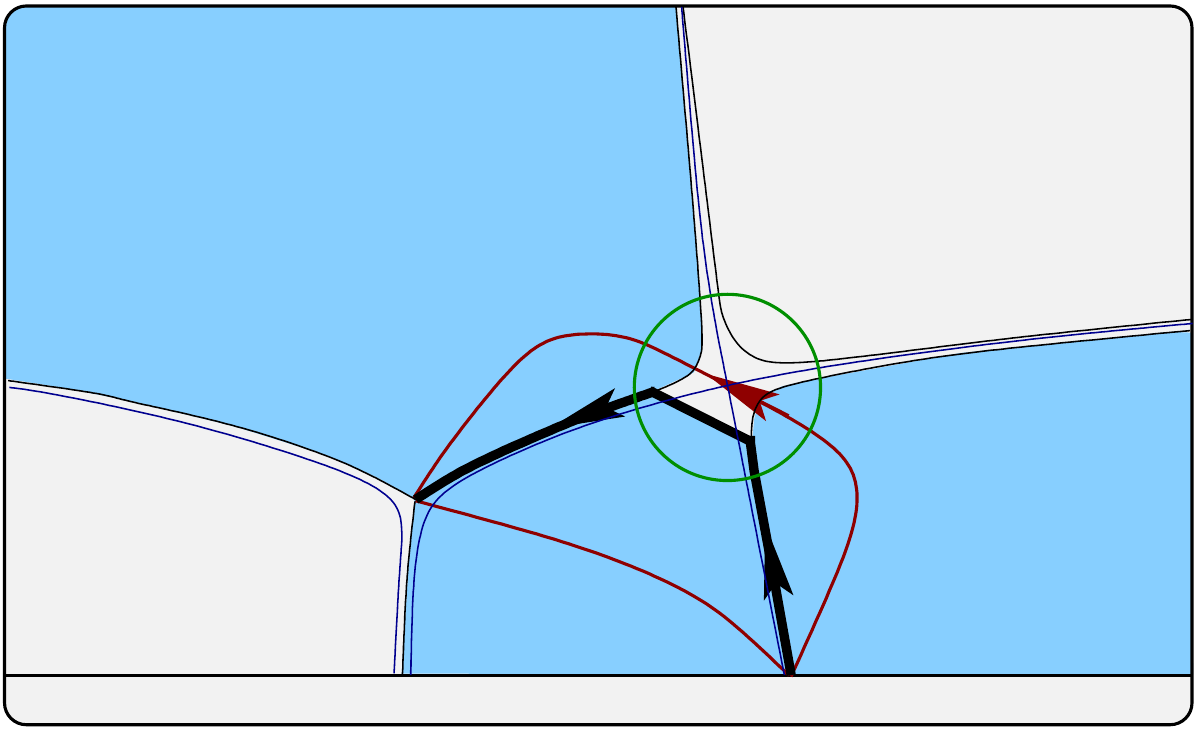}\hspace{0.1\textwidth}\resizebox{0.4\textwidth}{!}{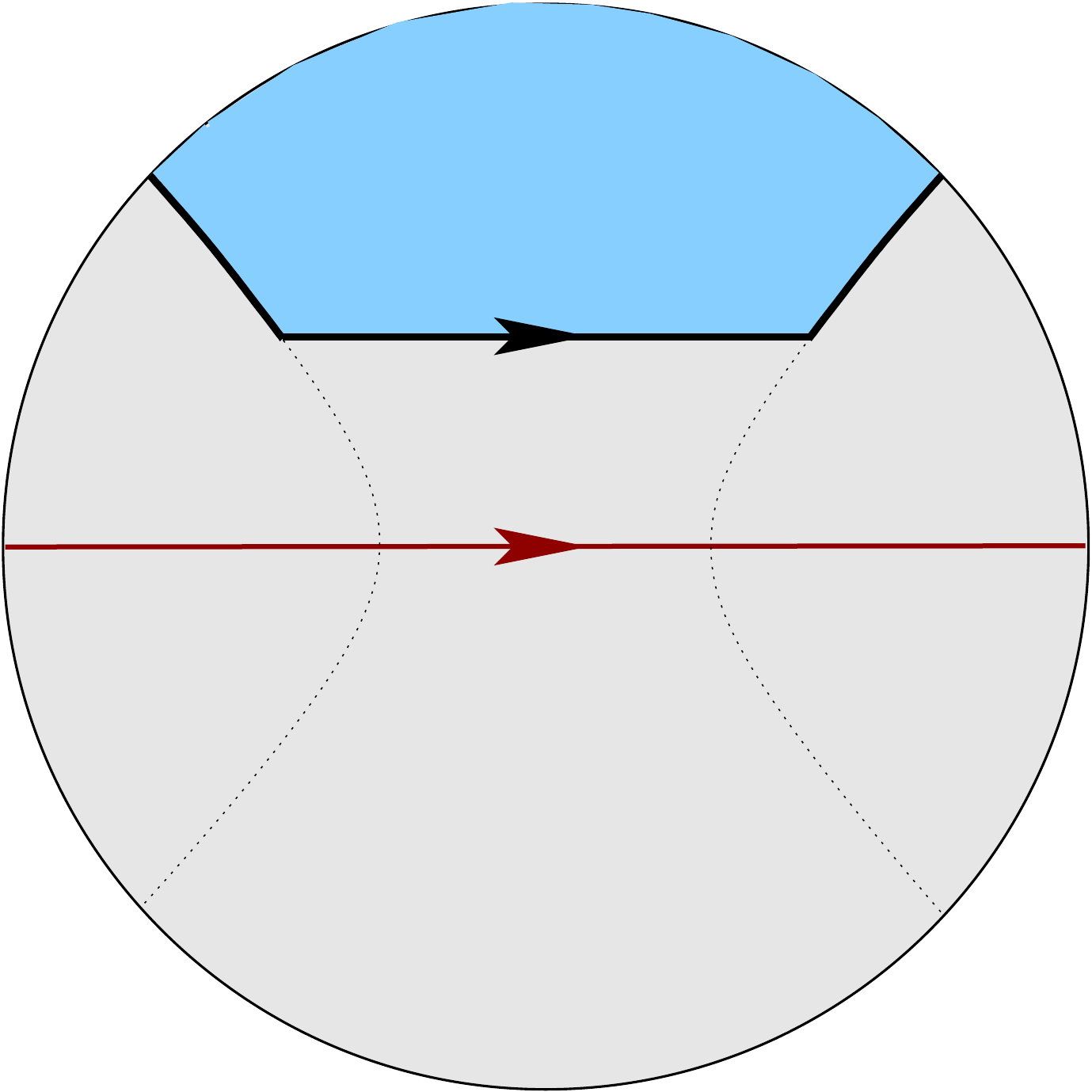}
\caption{[On the left] The main arc has split into two and we are modifying the function $h$ by introducing a small ``artificial'' main arc connecting the two pieces of the broken one.}
\label{ViolationPunished}
\caption{[On the right] The exact jumps of $Y$ in terms of the conformal coordinate $\zeta$ near the point $\eta$.}
\label{surgical}
\end{figure}

We then must re-define the function $h$ by introducing a small main arc within the local disk $\D$ according to Fig. \ref{ViolationPunished}

The difference between the two pictures on Fig. \ref{ViolationBreak} and Fig. \ref{ViolationPunished} is that we have changed the sign of $h$ within the region that has changed color; the result is a function that we still denote by $h(z)$, which is harmonic analytic on the upper complex plane minus the main arc (marked by the thick line).
In the sequel we will denote by $h(z)$ the function defined this way: it has the property that it is now {\bf analytic} in  a small neighborhood of $z=\eta$ and behaves
\bea
2i h(z) = -C(z-\eta)^2(1 + \mathcal O(z-\eta)) + 2iS,\\ 
C = C(x,t)\ ,\ \ \eta= \eta(x,t)\ ,\ \ S:=S(x,t)\ .
\eea
The arc made by the two lenses can be deformed to pass through the level-curve $\Re(h-S)=0$ and the zooming coordinate $\zeta$ can be defined as the conformal change of coordinate

\be
\zeta^2 = \le\{\begin{array}{cc}
\frac {2i}\varepsilon \le( h(z)-S\ri),  & z\in \D_1,\\[7pt]
-\frac {2i}\varepsilon\le( h(z) -S\ri), & z\in \D_0.
\end{array}
\ri.
\ee
We note that now we have $\Im (S)>0$ and hence it is convenient to parametrize it by 
\be
S:= 2 \vartheta + i 2\varkappa \varepsilon\ln \varepsilon,
\ee
where --contrary to the case of the right breaking curve-- {\bf negative}  values of $\varkappa$ correspond to the exploration into the genus--two region.
In Fig. \ref{surgical} we have shown the jumps and contours for the solution $Y(z)$ in the local disk $\D$. The model problem solves the usual jump $\Psi_+ = \begin{pmatrix} 0 & 1 \\ -1 &0\end{pmatrix}\Psi_-$.
\subsection{Outer parametrix}
\label{sectleftouter}
The situation from this point on is quite similar to the previous; the outer parametrix $\Psi_{_J}$ will be taken as defined by the same Riemann--Hilbert Problem \ref{KRHP}, with the only difference that the jump condition now runs on the piecewise-smooth arc. 
Note that the point $\eta$ (by our construction) lies to the right of the main arc; worth mentioning is also that this time we will have to use the formula for $\Psi_{_J}$ for the closest {\bf negative integer} to $\varkappa$. The formul\ae\ for $\Psi_{_J}$ are identical.
\subsection{Local parametrix}
\label{sectleftlocal}
Due to the different shape of the jumps within the local disk $\D$, the relationship between the present case and the previous ones are less immediate than for the model problem.
\begin{problem}
\label{Pproblem}
The local parametrix is a matrix--valued function $\mathcal P_{_J}(\zeta) = \mathcal P_{_J}(\zeta;\varkappa, \vartheta)$ of the zooming coordinate $\zeta$ satisfying the following {\bf mixed RHP} ($J\in -\N$).
\bea
\mathcal P_{_J +}(\zeta) &\&= \begin{bmatrix}
1 &  - \varepsilon^\varkappa {\rm e}^{-\zeta^2  - \frac i\varepsilon \vartheta}\\
0 & 1
\end{bmatrix} \mathcal P_{_J-}(\zeta)\ ,\qquad \zeta \in \R,\\
\mathcal P_{_J+}(\zeta) &\& = \begin{bmatrix} 0 & -1\\ 1 & 0 \end{bmatrix} \mathcal P_ {_J-} (\zeta)
\begin{bmatrix} 0 & 1\\ -1 & 0 \end{bmatrix}\ ,\ \ \zeta \in \hbox{\rm main arc},\label{mainarc}\\
\mathcal P_{_J}(\zeta) &\& = \zeta^{J\sigma_3} \mathcal O(1)\ ,\ \ \ \zeta\to 0,\\
\mathcal P_{_J}(\zeta) &\& = \1 + o(1)\ ,\ \ \ \zeta\in \pa \D.
\eea
\end{problem}
We anticipate that we will need this problem for negative integer $K$.
If we disregard the condition (\ref{mainarc}) then this problem is essentially (up to the triangularity of the jump) the same problem as for $\mathbf H_{_K}$; 
more precisely, let $K\in - \N$ and $\varkappa \in -\R_+$ and consider 
\be
\mathbf H_{_{(-J)}}(\zeta; -\varkappa,-\vartheta)
\ee
is the solution of  (\ref{Hjump}), (\ref{rowsing}), (\ref{boundarydecay}) with $K=-J$ and $\varkappa, \vartheta$ replaced by $-\varkappa, -\vartheta$. 
Then the solution of the RHP \ref{Pproblem} is 
\bea
\mathcal P_{_J} (\zeta; \varkappa,\vartheta) &\&:= \begin{bmatrix} 0& 1\\ 1 & 0 \end{bmatrix}\mathbf H_{_{(-J)}}(\zeta; -\varkappa,-\vartheta)  \begin{bmatrix} 0& 1\\ 1 & 0 \end{bmatrix} \ ,\qquad \zeta \in \D_1,\\
\mathcal P_{_J} (\zeta; \varkappa,\vartheta) &\&:= {\rm e}^{\frac {i\pi}2  \sigma_3}\mathbf H_{_{(-J)}}(\zeta; -\varkappa,-\vartheta) {\rm e}^{-\frac {i\pi}2  \sigma_3} \ ,\qquad \zeta \in \D_0.
\eea
It is promptly seen that $\mathcal P_{_J}$ solves the jump on the real axis and --on the boundary between $\D_0,\D_1$-- is satisfies
the condition (\ref{mainarc}).

It should be apparent now that the sequel of the analysis is quite parallel to the previous case; the parametrices $\Psi_{\varkappa}$ and $\mathcal P_{\varkappa}$ will be constructed by the same methods. 
The explicit formul\ae\ for the matrices $G_J, \wh G_J$ etc. have only minor cosmetic differences. 
\section{Corrections to the NLS solution}
\label{sectcorrect}

\bt
\label{mainthm}
The  {solution  of the NLS (\ref{FNLS}) defined by the scattering data $f_0(z)$}  behaves as 
\bea
q(x, t,\ve)  &\& = \le(-b(x,t){\rm e}^{2iK\psi(x,t)} -2\le(G_{K,\varkappa}(x,t,\ve) +  \wh  G_{K,\varkappa}(x,t,\ve)\ri)_{12} + 
\mathcal O(\sqrt{\varepsilon}) \ri) {\rm e}^{\frac {4i}\varepsilon g(\infty, x,t)} \label{914}
\eea
as $\ve\ra 0$.
Here $(x,t)$ are such that 
\be
\Im \le(S(x,t)\ri) = \frac {\varkappa }2  \varepsilon \ln \varepsilon\ ,\ \ \varkappa \in \R,
\ee
where 
$K =\max(0,\lfloor \varkappa\rfloor)$, $e^{i\psi} := \frac{\ov\l_0}{\l_0}$ (for $\l_0$ see (\ref{lambda0})), 
$G_{K,\varkappa}, \wh G_{K,\varkappa}$ have been defined in (\ref{Gkappas}). The term $(G_{K,\varkappa}+\wh G_{K,\varkappa})_{12}$ has the following behavior
\be
2(G_{K,\varkappa}+\wh G_{K,\varkappa})_{12} = 
\le\{\begin{array}{cc}
\ds  \mathcal O(\ve^{\frac 1 2 + K-\varkappa}), & \varkappa \in[K,K+\frac 1 2],\cr
 \ds - b(x,t)\Big[{\rm e}^{2iK\psi} - {\rm e}^{-2i(K+1)\psi} \Big] + \mathcal O (\ve^{\varkappa-K-\frac 12}), & \varkappa \in(K+\frac 1 2,K+1).
\end{array}\ri.\label{behave}
\ee
In particular the correction is of the order $O(1)$ one as $\ve\ra 0$ only for $\varkappa\in \frac 1 2 + \N$.
\et
The proof can be found in Appendix \ref{proofs}.
\br
Considering $\varkappa$ as a parameter in a direction transversal to the breaking curve, we notice that  centers of adjacent ripples  are at  unit  $\varkappa$-distance from each other. On the other hand, according to (\ref{behave}), the ``support'' of each ripple is of the order $O(\frac{1}{\ln \varepsilon})$ in terms of the exploration parameter $\varkappa$. Since the order of $\varkappa$ is $\mathcal O(\varepsilon \ln \varepsilon)$, this means that the support of each ripple is of order $\mathcal O (\varepsilon)$, exactly as in the bulk of the genus $2$ region. However their {\em spacing} is of the same order as $\varkappa$, so that they look like ranges of mountains/valleys separated by  vast flatlands (see Fig. \ref{3dpic}).
\er
While Theorem \ref{mainthm} contains the substantial qualitative information about the relevant features of the behavior, 
the reader may wish for a slightly more concrete formula. It is possible to write the correction term in full detail 
and --for example-- use some computer algebra software to achieve the plot in Fig. \ref{3dpic}. We restate the theorem with some more explicit expressions.
\bt
\label{mainthm2}
The  {solution  of the NLS (\ref{FNLS}) defined by the scattering data $f_0(z)$}  behaves as 
\be
q(x,t) = -b(x,t){\rm e}^{\frac {4ig(\infty)}{\ve} +2iK\psi} \le(1 - \frac {2{\rm e}^{-2iK\psi}} { b(x,t) \Delta_K} \bigg\{
|u_{_K}|^2 \chi_{_K}  + 
u_{_K}\, {\B_{_K,1}}^2 
+\ov u_{_K}\,  \ov{\B_{_K,2}}^2 
\bigg\} + \mathcal O(\sqrt{\ve}) \ri)~,
\ee 
where
\bea\label{BK1,2}
\B_{K,1} = \frac {\tau^K {\rm e}^{iK\psi}}{\sqrt{1 + \l_0^2}}, 
&&
\B_{K,2} = \frac {\tau^K {\rm e}^{-iK\psi}\l_0}{\sqrt{1 + \l_0^2}},
\\
\label{tau}
\t:=|\t|e^{i\phi}=\frac{\l_0^2(1+|\l_0|^2)}{ib\Im (\l_0)(1+\l_0^2)^2}, 
&&
\eea
\be\label{uK}
u_{_K} =  -i \frac {K!  }{\sqrt{\pi}2^{K+1}  C^{K+\frac 1 2}} \exp\le(i\frac {\vartheta} {\varepsilon}\ri) \varepsilon^{\frac 1 2 -\varkappa + K }~,
\ee
%
\bea
\Delta_K 
 =1+ 4 \Re\le( 
\frac{u_{_K}  \l_0 \tau^{2K}}{b(1 + \l_0^2)^2}
\ri) + 
\frac {4 |u_{_K}|^2 |\l_0|^4 |\tau|^{4k}  }{b^2 \le|1+\l_0^2 \ri|^4} \le( 1+ \frac {|1+\l_0^2|^2}{4\Im(\l_0)^2}
\ri),
 \eea
\be\label{chi}
\chi_K=\frac{e^{2iK\psi}}{b|1+\l_0^2|^2}\le(\frac{2\ov\l_0^2\tau^{2K} }{1+\ov\l_0^2}+\frac{2|\l_0|^4 \ov \tau^{2K}}{1+\l_0^2}-\frac{ib\ov\l_0|1+\l_0^2|}{\Im \eta}\ri).
\ee
\et
The proof can be found in App. \ref{proofs}.

\subsection{Discussion}
\label{secdiscussion}

\paragraph {\bf Universality.} The correction depends only on $\Psi_K$, $\eta$ and the constant $C$ that appears in $u_K$; in particular, 
$\Psi_K$ depends only on the branch-point $\alpha$ and the integer $K$ which counts the number of first ripples. 
Thus, universality here is taken to mean that no {details of $f_0(z)$, that defines a particular solution to the NLS (\ref{FNLS}),} 
enter into the formula.

Since the gradient of  $\Im S(x,t)$ does not vanish, the natural scale of modulation of $q(x,t)$ in the direction transversal to the breaking curve is $\varepsilon\ln \frac 1 \varepsilon  \|\nabla \Im S\|$ and $\varkappa$ serves as exploration parameter in this scale.

{Consider the case $\varkappa= K+\frac 1 2  \in \frac 1 2 +\N$, i.e. on the ``crest'' of the $K$-th ripple; 
for this value of the parameter the correction to $q(x,t)$ is of order $\mathcal O(1)$. In this case $u_{K}$ is a fast oscillatory function: if we fix a small $\varepsilon$ then the correction term to the amplitude has a slow modulation in $(x,t)$ due to the dependence of $C,\eta, \alpha$ etc. {\em and} a fast change along the direction parallel to the breaking curve due to the fast oscillation of $u_{_K}$. }

\paragraph {\bf Large time limit of the profile.} 

According to \cite{TVZ2}, \cite{TVZ3}, together with (\ref{lambexp}), we have the following  asymptotics along the right breaking curve as $t\ra\infty$:
\bea\label{largetass}
&\a=a+ib \sim \m_++ie^{\frac{2\pi(\m_+-\m_-)t}{w'(\m_+)}},  \\
&\eta\sim\m_-+i\frac{w'(\m_-)}{4t}, \\
&\l_0=\l(\eta)\sim \frac{-2(\m_+-\m_-)}{b}+i\frac{w'(\m_-)}{2tb},\\
&\t\sim -2i\frac{e^{i\psi}t}{w'(\m_-)}.\ ,\qquad w(x) := \Im \le(f_{0}(x)_+\ri) \sign(x-\mu_+)
\eea
Then 
\be\label{largetBK2}
\ov\B^2_{K1}\sim e^{2iK\psi}i^{2K}\le(\frac{2t}{w'(\m_-)}\ri)^{2K},
\ee
whereas $\det[\B'_K,\B_K]$ and $\B_{K1}$ are decaying exponentially as $t\ra\infty$. Thus
\be
r_K=\Delta_K\sim 1+|u_K|^2\le(\frac{2t}{w'(\m_-)}\ri)^{4K+2}
\ee
and, according to (\ref{Cass}) and (\ref{uK}),
\be
u_K\sim e^{-\frac{i\pi}{4}}e^{-i\frac{\theta}{\varepsilon}}\frac{\varepsilon^{K+\hf-\varkappa}i^K K!}{\sqrt{\pi}2^{3K+2}t^{K+\hf}}~.
\ee
{In order to gain some insight on the profile and relative height of the ripples for large time we take $\varkappa$ a half--integer 
(similar consideration hold in a double-scaling approach as long as the expression $\varepsilon^{K+\hf-\varkappa}t^{K+\hf}$ approaches infinity)}. Then we obtain ($\varkappa = K+\frac 1 2$)
\be\label{rippleass}
\big(G_{K,K+\frac 12}-\ov G_{K,K+\frac 12}\big)_{12} \sim  -\sqrt{\pi}e^{-\frac{i\pi}{4}}i^Ke^{2iK\psi}e^{i\frac{\theta}{\varepsilon}}\frac{2^Kw'(\m_-)^{2K+2}}{K! \,\,t^{K+\frac{3}{2}}}~.
\ee
In the particular example, considered in the paper for numerical computations, $w'(\m_-)=\hf\pi$. Notice that the ripples has algebraic decay as $t\ra\infty$ whereas
the amplitude of the corresponding genus zero solution decays exponentially fast. Thus, $F_{ripple}$ is the main contribution to the solution along the ripples
for large $t$. Expression (\ref{rippleass})  also shows that the rate of decay for large $t$ increases with the number $K$; however, it may be offset by the $K!$
growth of the coefficient in front of $t^{-K-\frac{3}{2}}$. That is consistent with $\mathcal O(t^{-\hf})$ magnitude of the solution in the  bulk of the genus two region,
see \cite{TVZ2}.

\begin{remark}
Similar results are true for transition through any breaking curve separating regions of genus $2N$ and $2N+2$. 
\end{remark}

\br
This type of scales are to be expected from the similarities with random matrix models \cite{Eynard:2006p10}.
\er
\appendix
\renewcommand{\theequation}{\Alph{section}.\arabic{equation}}
\section{Explicit formul\ae}
\label{appexplicit}

Using (\ref{model0}), the expression for $\Phi_K(z)$ can be written as
\be\label{PhiKz}
\Phi_{_K}(z) =\le( \frac{\ov \lambda_0}{\lambda_0} \ri)^{K\sigma_3}\Psi_{_0} (z)
\left[(z-\eta)D(z) \right]^{K\sigma_3}~.
\ee 
To express it in terms of $\l$, we calculate 
\be\label{z-etaD}
(z-\eta)D(z)=\frac{b(\l-\ov\l_0)(1+\l\l_0)^2}{2\l\l_0(1+\l\ov\l_0)},
\ee
so that (\ref{PhiKz}) becomes
\be\label{PhiKlam}
\Phi_{_K} =\frac{1}{\sqrt{\l^2+1}}\le( \frac{\ov \lambda_0}{\lambda_0} \ri)^{K\sigma_3}
\begin{bmatrix}
\l & -1\\
1 & \l 
\end{bmatrix}
\left( \frac{b(\l-\ov\l_0)(1+\l\l_0)^2}{2\l\l_0(1+\l\ov\l_0)}\right)^{K\sigma_3}~.
\ee

%

The quantities involved in (\ref{Reality}) require the computation of  $\Phi_{_K}'(z)$, which is given by
\be\label{Phi'Kz}
\Phi'_{_K}(z) =\le( \frac{\ov \lambda_0}{\lambda_0} \ri)^{K\sigma_3}\left( \frac{z-\a}{z-\ov\a}\right)^{\frac{\s_2}{4}}
\left[\frac{ib\s_2}{2R^2(z)}+K\s_3\left(\ln(z-\eta)+\ln D(z)\right)'\right] 
\left((z-\eta)D(z) \right)^{K\sigma_3}~.
\ee
The expression in the square brackets is
\be\label{M}
\frac{2\l^2}{b(1+\l^2)}\left[ \frac{i}{1+\l^2}\s_2+K\nu(\l)\s_3\right],
\ee
where
\be
\nu(\l)=\left(\frac{2\l_0}{1+\l\l_0}-\frac{\ov\l_0}{1+\l\ov\l_0}+\frac{1}{\l-\ov\l_0}-\frac{1}{\l} \right)~. 
\ee
Thus,
\be\label{Phi'K}
\Phi'_{_K}(z)=
\frac{2\l^2}{b(1+\l^2)^\frac{3}{2}}\le( \frac{\ov \lambda_0}{\lambda_0} \ri)^{K\sigma_3}
\begin{bmatrix}
K\l\nu(\l)+\frac{1}{1+\l^2} & K\nu(\l)+\frac{\l}{1+\l^2}\\
K\nu(\l)-\frac{\l}{1+\l^2} & -K\l\nu(\l)+\frac{1}{1+\l^2}
\end{bmatrix}
\left( \frac{b(\l-\ov\l_0)(1+\l\l_0)^2}{2\l\l_0(1+\l\ov\l_0)}\right)^{K\sigma_3}~.
\ee
%
%
\section{Proofs}
\label{proofs}
\subsection{Proof of Prop. \ref{propcont}.}
Equation (\ref{DeltaGK}) follows immediately from (\ref{conteq}) by comparing the behavior of both sides of the equation  as $z\to \infty$.
To verify this we recall that the matrices $G_{K,\varkappa}, \wh G_{K,\varkappa}, F_{K+1,\varkappa}, \wh F_{K+1,\varkappa}, C_K, \wh C_K$ are uniquely characterized by the equations (\ref{bound1}, \ref{bound2}) that can be written as
\bea
\le\{ \begin{array}{c}
\le(\1 - \frac {G_{K,\varkappa}}{z-\eta} - \frac {\wh G_{K,\varkappa}}{z-\ov \eta}\ri) \le[\A_K(z),\B_K(z)\ri] \le[
\begin{array}{cc}
1  &0 \\
\ds  \frac{u_K}{z-\eta}  & 1
\end{array}
\ri]= \mathcal O(1) \ \ \ \ z\to \eta\cr
\le(\1 - \frac {G_{K,\varkappa}}{z-\eta} - \frac {\wh G_{K,\varkappa}}{z-\ov \eta}\ri) \le[\wh \A_K(z),\wh \B_K(z)\ri] \le[
\begin{array}{cc}
1  &\ds  \frac{  \wh u_K }{z- \ov \eta} \\
0  & 1
\end{array}
\ri] = \mathcal O(1),\ \ \ \ z\to \ov\eta, 
\end{array}\ri .\\
\le\{\begin{array}{c}
\le(\1 - \frac {F_{K+1,\varkappa}}{z-\eta} - \frac {\wh F_{K+1,\varkappa}}{z-\ov \eta}\ri) \le[\A_{K+1}(z),\B_{K+1}(z)\ri] \le[
\begin{array}{cc}
1  &\ds  \frac{\ell_{K+1}}{z-\eta} \\
0  & 1
\end{array}
\ri]= \mathcal O(1)\ \ z\to \eta\cr
\le(\1 - \frac {F_{K+1,\varkappa}}{z-\eta} - \frac {\wh F_{K+1,\varkappa}}{z-\ov \eta}\ri) \le[\wh \A_{K+1}(z),\wh \B_{K+1}(z)\ri] \le[
\begin{array}{cc}
1  & 0 \\
\ds  \frac{  \wh \ell_{K+1} }{z- \ov \eta}   & 1
\end{array}
\ri] = \mathcal O(1),\ \ z\to \ov\eta,
\end{array}\ri.\label{Fk+}
\eea
\be
\ell_K :=  
 \frac {2i\pi  {\rm e}^{-i\frac \vartheta {\varepsilon}} \varepsilon^{\varkappa - K + \frac  1 2}} {\mu_{K-1} C^{-K+\frac 1 2}}\ ,\qquad
\wh \ell_K := 
 \frac {2i\pi {\rm e}^{i\frac \vartheta {\varepsilon}} \varepsilon^{\varkappa - K + \frac  1 2}} {\mu_{K-1} \wh C^{-K+\frac 1 2}}\label{ellk}
\ee 
\bea
\le\{\begin{array}{c}
\le(\1 - \frac {C_K}{z-\eta} - \frac {\wh C_K}{z-\ov \eta}\ri) \le[ \A_K(z), \B_K(z)\ri] = \le[ \A_{K+1}(z), \B_{K+1}(z)\ri](z-\eta)^{-\sigma_3} \\
\le(\1 - \frac {C_K}{z-\eta} - \frac {\wh C_K}{z-\ov \eta}\ri) \le[ \wh \A_K(z), \wh\B_K(z)\ri] = \le[\wh \A_{K+1}(z), \wh\B_{K+1}(z)\ri](z-\ov \eta)^{\sigma_3}. 
\end{array}\ri.
\label{Cks}\eea
We note that  from (\ref{uk},\ref{ellk}) it follows
\be
u_K \ell_{K+1} = 1 = \wh u_K \wh \ell_{K+1}.
\ee
So the check consists in verifying that the matrix 
\be
M(z):=  \le(\1 - \frac {F_{K+1,\varkappa}}{z-\eta} - \frac {\wh F_{K+1,\varkappa}}{z-\ov \eta} \ri) \le(\1 - \frac{C_K}{z-\eta} - \frac {\wh C_K}{z-\ov \eta}\ri)
\ee
solves 
\be
\le\{ \begin{array}{c}
M(z) \le[\A_K(z),\B_K(z)\ri] \le[
\begin{array}{cc}
1  &0 \\
\ds  \frac{u_K}{z-\eta}  & 1
\end{array}
\ri]= \mathcal O(1)\cr
M(z) \le[\wh \A_K(z),\wh \B_K(z)\ri] \le[
\begin{array}{cc}
1  &\ds  \frac{  \wh u_K }{z- \ov \eta} \\
0  & 1
\end{array}
\ri] = \mathcal O(1). 
\end{array}\ri .
\ee
Indeed we have
\bea
&& M(z) \le[\A_K(z),\B_K(z)\ri] \le[
\begin{array}{cc}
1  &0 \\
\ds  \frac{u_K}{z-\eta}  & 1
\end{array}
\ri]=\cr
&&= \le(\1 - \frac {F_{K+1,\varkappa}}{z-\eta} - \frac {\wh F_{K+1,\varkappa}}{z-\ov \eta} \ri)\  \le(\1 - \frac{C_K}{z-\eta} - \frac {\wh C_K}{z-\ov \eta}\ri)\le[\A_K(z),\B_K(z)\ri] \le[
\begin{array}{cc}
1  &0 \\
\ds  \frac{u_K}{z-\eta}  & 1
\end{array}
\ri]=
\cr
&& \hbox{[From (\ref{Cks})]}= \le(\1 - \frac {F_{K+1,\varkappa}}{z-\eta} - \frac {\wh F_{K+1,\varkappa}}{z-\ov \eta} \ri)
\le[\A_{K+1}(z),\B_{K+1}(z)\ri]
(z-\eta)^{-\sigma_3}\le[
\begin{array}{cc}
1  &0 \\
\ds  \frac{u_K}{z-\eta}  & 1
\end{array}
\ri] =\cr
&& \hbox{[From (\ref{Fk+})]}
= \mathcal O(1) 
\le[
\begin{array}{cc}
1  &\ds - \frac{\ell_{K+1}}{z-\eta} \\
0  & 1
\end{array}
\ri]
(z-\eta)^{-\sigma_3} \le[
\begin{array}{cc}
1  &0 \\
\ds  \frac{u_K}{z-\eta}  & 1
\end{array}
\ri]  = \cr
&& = \mathcal O(1) \le[
\begin{array}{cc}
\frac {1-\ell_{K+1} u_K} {z-\eta}  & -\ell_{K+1}\\ u_K & (z-\eta).
\end{array}
\ri]
\eea
This last expression is analytic because $\ell_{K+1} u_K =1$. This proves the identity (\ref{conteq}). One should repeat the verification near $\ov \eta$ but it is completely parallel. This proves the continuity.
\QED
\subsection{Proof of Thm. \ref{thmpsikappa}.}
The existence of the solution will be implied by the formula and by noticing that all denominators are strictly positive.
The symmetries (\ref{Gsym1}) are obtained from the symmetries (\ref{symPhi}), (\ref{symPhi'}).

The conditions that determine uniquely $G_{K,\varkappa}, \wh G_{K,\varkappa}, F_{K,\varkappa}, \wh F_{K,\varkappa}$ are to be read off (\ref{bound1},\ref{bound2}):  we will show the details only for the pair $G_{K,\varkappa}, \wh G_{K,\varkappa}$, the computation for $F_{K,\varkappa}, \wh F_{K,\varkappa} $ being completely similar.

%
%
\paragraph{Case $K<\varkappa < K+\frac 1 2$.}
The conditions (\ref{bound1}, \ref{bound2}) translate into
\bea
\le(\1 - \frac {G_{K,\varkappa}}{z-\eta} - \frac {\wh G_{K,\varkappa}}{z-\ov \eta}\ri) \le[\A_K(z),\B_K(z)\ri] (z-\eta)^{-K\sigma_3}\le[
\begin{array}{cc}
1  &0 \\
\ds \frac{ 
 \mu_K}{2i\pi \zeta^{2K+1}  {\rm e}^{-i\frac \vartheta {\varepsilon}} \varepsilon^\varkappa}  & 1
\end{array}
\ri]  (z-\eta)^{K\sigma_3} = \mathcal O(1),\\
\le(\1 - \frac {G_{K,\varkappa}}{z-\eta} - \frac {\wh G_{K,\varkappa}}{z-\ov \eta}\ri) \le[\wh \A_K(z),\wh \B_K(z)\ri] (z-\ov\eta)^{K\sigma_3}\le[
\begin{array}{cc}
1  &\ds \frac{  \mu_K}{2i\pi \wh \zeta^{2K+1} {\rm e}^{i\frac \vartheta {\varepsilon}} \varepsilon^\varkappa} \\
0  & 1
\end{array}
\ri]  (z-\ov \eta)^{-K\sigma_3} = \mathcal O(1).
\eea
This can be rewritten
\bea
\le(\1 - \frac {G_{K,\varkappa}}{z-\eta} - \frac {\wh G_{K,\varkappa}}{z-\ov \eta}\ri) \le[\A_K(z),\B_K(z)\ri] \le[
\begin{array}{cc}
1  &0 \\
\ds \le (\frac{z-\eta}{\zeta \sqrt {\varepsilon}}\ri)^{2K} \frac{
 \mu_K}{2i\pi \zeta  {\rm e}^{-i\frac \vartheta {\varepsilon}} \varepsilon^{\varkappa-K}}  & 1
\end{array}
\ri]= \mathcal O(1),\\
\le(\1 - \frac {G_{K,\varkappa}}{z-\eta} - \frac {\wh G_{K,\varkappa}}{z-\ov \eta}\ri) \le[\wh \A_K(z),\wh \B_K(z)\ri] \le[
\begin{array}{cc}
1  &\ds  \le (\frac{z-\ov \eta}{\wh \zeta \sqrt {\varepsilon}}\ri)^{2K}\frac{  \mu_K}{2i\pi \wh \zeta  {\rm e}^{i\frac \vartheta {\varepsilon}} \varepsilon^{\varkappa-K}} \\
0  & 1
\end{array}
\ri] = \mathcal O(1).
\eea
From  (\ref{zoomcoord}) it follows that 
\be
\sqrt \varepsilon \zeta = \sqrt {C} (z-\eta) (1 + \mathcal O(z-\eta))\ ,\qquad
\sqrt \varepsilon \wh \zeta = \sqrt {\wh C} (z-\ov\eta) (1 + \mathcal O(z-\ov \eta)).
\ee
In particular, we have 
\be
\lim_{z\to \eta} \frac {z-\eta}{\zeta \sqrt{\varepsilon}} = \frac 1{\sqrt C}\ , \qquad
\lim_{z\to \ov \eta} \frac {z-\ov \eta}{\wh \zeta \sqrt{\varepsilon}} = \frac 1{\sqrt {\wh C}}
\ee
and that one can replace the above condition by the following 
\bea
\le(\1 - \frac {G_{K,\varkappa}}{z-\eta} - \frac {\wh G_{K,\varkappa}}{z-\ov \eta}\ri) \le[\A_K(z),\B_K(z)\ri] \le[
\begin{array}{cc}
1  &0 \\
\ds  \frac{u_K}{z-\eta}  & 1
\end{array}
\ri]= \mathcal O(1),\cr
\le(\1 - \frac {G_{K,\varkappa}}{z-\eta} - \frac {\wh G_{K,\varkappa}}{z-\ov \eta}\ri) \le[\wh \A_K(z),\wh \B_K(z)\ri] \le[
\begin{array}{cc}
1  &\ds  \frac{  \wh u_K }{z- \ov \eta} \\
0  & 1
\end{array}
\ri] = \mathcal O(1),\label{Gks}\\
u_K:=
\frac {\mu_K C^{-K-\frac 12}}{2i\pi {\rm e}^{-i\frac \vartheta {\varepsilon}} \varepsilon^{\varkappa - K - \frac 1 2 }}\ ,\qquad
\wh u_K:=
 \frac {\mu_K\wh C^{-K-\frac 1 2} }{2i\pi  {\rm e}^{i\frac \vartheta {\varepsilon}} \varepsilon^{\varkappa - K - \frac 1 2 }}\ ,\ \ \mu_K:= \frac {K!\sqrt{\pi}}{2^K}.
\eea
We remark that 
\be
\wh u_K = - \ov u_K\ .\label{716}
\ee
This implies the following system of equations 
\bea
\le\{
\begin{array}{c}
G_{K,\varkappa} \B_K =0\\[10pt]
\wh G_{K,\varkappa}\wh \A_K =0\\[10pt]
\ds -G_{K,\varkappa} \A_K + u_K \le( \B_K - G _K\B'_K -\frac{ \wh G_{K,\varkappa} \B_K}{\eta-\ov \eta}\ri) =0\\[10pt]
\ds -\wh G_{K,\varkappa}\wh   \B_K + \wh u_K \le( \wh \A_K - \frac{  G_{K,\varkappa} \wh \A_K}{\ov \eta- \eta} - \wh G_{K,\varkappa} \wh \A'_K\ri) =0~,
\end{array}\ri.\label{Gsyst}
\eea
where --for brevity-- $\A_K := \A_K(\eta), \A'_K:= \A'_K(\eta)$ etc.
The solution is 
\bea
G_{K,\varkappa} (\bullet)= \frac {u_K \det [ \bullet, \B_K] {(\eta-\ov \eta)^2} }{ \le(1+ u_K\det[\B_K', \B_K]\ri) \le(1 - \wh u_K \det[\wh \A_K', \wh \A_K]\ri){(\eta-\ov \eta)^2}   +  {u_K \wh u_K \det[\wh \A_K , \B_K ]^2} } \times\\
\times
\le(\le(1 - \wh u_K \det[\wh \A_K' , \wh \A_K]\ri) \B_K + \frac {\wh u_K \det[\B_K, \wh \A_K]}{(\eta - \ov \eta)} \wh \A_K\ri)~,
\\
%
%
%
%
\wh G_{K,\varkappa}( \bullet)=  \frac { -  \wh u_K \det [ \bullet, \wh \A_K](\ov\eta- \eta)^2 }{ {\le(1 +  u_K \det[ \B'_K,  \B_K]\ri)}\le(1- \wh u_K\det[\wh \A_K', \wh \A_K]\ri)(\ov\eta- \eta)^2 +  
{\wh u_K  u_K \det[\wh \A_K,  \B_K]^2} } \times\\
\times 
\le({\le(1 +  u_K \det[ \B'_K,  \B_K]\ri)}\wh \A_K - \frac { u_K \det[\wh \A_K,  \B_K]}{(\ov \eta -  \eta)}  \B_K\ri)~,
\eea
where we have used that 
\be
\det[\A_K(z), \B_K(z)] \equiv \det[\wh \A_K(z),\wh \B_K(z)]\equiv 1.
\ee

Using the reality relations  (\ref{Reality}) and (\ref{716}) we can rewrite them in the slightly simpler form which is in the statement of the theorem. 
\paragraph{Case $K-\frac 1 2< \varkappa <K$.} 
The derivation
is completely analog. In particular we will see below that the formul\ae\  for $F_{K,\varkappa}, \wh F_{K,\varkappa}$ are not necessary for our purposes if we have those for $G_{K,\varkappa}, \wh G_{K,\varkappa}$.
\QED

\subsection{Proof of Thm. \ref{mainthm}.}
According to (\ref{nlssolfull}), the solution to NLS is recovered from the solution $\Gamma(z)$ of the {\em original} RHP as 
\be
q(x,t,\ve) = -2\lim_{z\to\infty} z \Gamma_{12}(z).
\ee
Away from the cuts the approximation is such that 
\be
{\rm e}^{-\frac{2 i}\varepsilon g(\infty)\sigma_3} \Gamma(z) {\rm e}^{\frac {2i}\varepsilon g(z) \sigma_3} = \mathcal E(z) \Psi_\varkappa(z)~,\label{92}
\ee
where $\mathcal E(z)$ is the error term matrix which is (uniformly) $\mathcal E(z) = \1 + \mathcal O(\sqrt {\varepsilon})$ and, {according to (\ref{gform}),} 
\be
g(z;x,t) = g(\infty;x,t) + \mathcal O(z^{-1})\ ,\ \ \ \phi\in \R.
\ee
Thus we have 
\be
q(x,t,\ve) = -2 {\rm e}^{\frac {4i}\varepsilon g(\infty;x,t)} \lim_{z\to\infty}z \Psi_{\varkappa,12}(z).
\ee

According to Thm. \ref{thmpsikappa},  Cor. \ref{corpsikappa} and Proposition \ref{propPsiK}, 
\be \label{Psivark}
\Psi_{\varkappa}(z)=T_{K, \varkappa}(z)\left( \frac{\ov\l_0}{\l(z)}\right)^{K\s_3} \Psi_0(z) \left( D(z)\right)^{K\s_3}\qquad K =\max(0,\lfloor \varkappa \rfloor).
\ee
 Then
\bea
-2\lim_{z\to \infty} z [\Psi_\k]_{12}  &\& =-2\le(\frac{\ov \lambda_0}{\lambda_0} \ri)^{2K} \lim_{z\to \infty}z[\Psi_0]_{12}+2(G_{K,\varkappa} + \wh G_{K, \varkappa})_{12} =\cr
&\& = -b(x,t) {\rm e}^{2iK\psi}
+2(G_{K,\varkappa} - \wh G_{K, \varkappa})_{12}\cr
&\& {\rm e}^{i\psi}: =\frac {\ov \l_0}{\l_0}.
\eea
where $\alpha(x,t) = a(x,t)+i b(x,t)$. 
This is the leading order in the approximation: { the correction to the amplitude $b(x,t)$ of 
the solution $q(x,t,\ve)$} is purely a phase-shift in units of $\psi$, 
{ where $ {\rm e}^{i \psi} =\frac{\ov \lambda_0}{\lambda_0} $;
the integer $K$  counts} the number of ``ripples'' as we explore the genus two region. So, finally,
\bea
q(x,t,\ve) &\&= -2{\rm e}^{\frac {4i}\varepsilon g(\infty)} \lim_{z\to \infty} z\le(\mathcal E(z)\Psi_{\varkappa}(z)\ri)_{12} = \cr
&\& =\le(-b(x,t){\rm e}^{2iK\psi} + 2
(G_{K,\varkappa} + \wh G_{K,\varkappa})_{12} + \mathcal O(\sqrt{\varepsilon})\ri) {\rm e}^{\frac {2i}\varepsilon g(\infty)}
\qquad  K\leq \varkappa <K+1
\eea
This proves formula (\ref{914}).
To verify the assertion regarding the different behaviors (\ref{behave}) we recall here equation (\ref{DeltaGK}) 
\be
G_{K,\varkappa} + \wh G_{K,\varkappa}  = F_{K+1,\varkappa} +\wh F_{K+1,\varkappa} + C_K + \wh C_K\ .
\ee
We know that the matrices $ F_{K+1,\varkappa}, \ \wh F_{K+1,\varkappa} $ both are of order $\mathcal O(\ve^{\varkappa - K - \frac 12 })$ (by inspection of (\ref{Fkappas}) and (\ref{uk})).
On the other hand
\be
C_K + \wh C_K = \lim_{z\to \infty} z \le(\Psi_K \Psi_{K+1}^{-1}\ri) = \frac {- b(x,t)}{2} \le[{\rm e}^{i2K\psi} - {\rm e}^{-2(K+1)\psi}\ri]~.
\label{DeltaCK}
\ee
And this concludes the proof. \QED
%
%
%
%
\subsection{Proof of Thm. \ref{mainthm2}.}
 Looking at the formul\ae\ (\ref{Gkappas}) and recalling formul\ae\ (\ref{Reality}) and  $\wh u_K = -\ov {u_K}$ we find
\bea\label{GK+ovGK12}
(G_{K,\varkappa}+\wh G_{K,\varkappa})_{12} &\& = \frac {-1}{\Delta_K}
\bigg\{
|u_{_K}|^2 \chi_{_K}  + 
u_{_K}\, {\B_{_K,1}}^2 
+\ov u_{_K}\,  \ov{\B_{_K,2}}^2 
\bigg\},\\
\chi_{_K} &\& :=\le(\ov{ \det[\B_{_K}' , \B_{_K}]} {\B_{_K,1}}^2  +\det[\B_{_K}',\B_{_K}] \ov{\B_{_K,2}}^2  -  2 \frac{\|\B_{_K}\|^2  \ov \B_{_K,2} \B_{_K,1}}{\ov \eta -\eta}\ri),\label{chimarco}\\
\Delta_K&\& :=   \Big |1 +  u_{_K} \det[ \B'_{_K},  \B_{_K}]\Big |^2  +  
\frac {| u_{_K} |^2 \| \B_{_K}\|^4} {4(\Im(\eta))^2} >0.
\eea
At this point one needs to use (\ref{PhiKlam}) (which defines the column vectors $\A_K, \B_K$) and  (\ref{Phi'K}) (which gives $\A_K', \B_K'$), which produce (\ref{BK1,2}). Plugging into the various object we obtain 
\bea
\label{det}
\det[\B_K', \B_K]=\frac{2\l_0^2}{b(1+\l_0^2)^2}
{\tau ^{2K}}
&& 
\parallel \B_K\parallel^2=\frac{1+|\l_0|^2 }{|1+\l_0^2|}|\t|^{2K}
\eea
The rest is simply an algebraic manipulation which is as straightforward as uninteresting.
\QED

\bibliographystyle{unsrt}
\bibliography{/Users/bertola/Documents/Papers/BibDeskLibrary.bib}
\end{document}